\documentclass[aps,prd,twocolumn,showpacs,superscriptaddress,groupedaddress,eqsecnum,nofootinbib,floatfix]{revtex4-1}

\usepackage{amsmath}
\usepackage{amsfonts}
\usepackage{amssymb}	
\usepackage{graphicx}
\usepackage{bm}
\usepackage{color}
\usepackage{commath}
\allowdisplaybreaks

\def\TEOBResumS{\texttt{TEOBResumS}}

\begin{document}
\title{Strong-field scattering of two black holes: Numerical Relativity meets Post-Minkowskian gravity}

\author{Thibault \surname{Damour}${}^{1}$}
\author{Piero \surname{Rettegno}${}^{2,3}$}

\affiliation{${}^1$ Institut des Hautes Etudes Scientifiques, 91440 Bures-sur-Yvette, France}
\affiliation{${}^2$ INFN Sezione di Torino, Via P. Giuria 1, 10125 Torino, Italy}
\affiliation{${}^3$ Dipartimento di Fisica, Universit\`a di Torino, via P. Giuria 1, 10125 Torino, Italy}

\begin{abstract}
	We compare numerical relativity (NR) data on the scattering of equal-mass, non-spinning binary black holes to 
	various analytical predictions based on post-Minkowskian (PM) results.	
	While the usual sequence of PM-expanded scattering angles shows a rather poor convergence towards NR data, 
	we find that a reformulation of PM information in terms of Effective-One-Body radial potentials
	leads to remarkable agreement with NR data, especially when using the radiation-reacted 4PM information.
	Using Firsov's inversion formula we directly extract, for the first time, from NR simulations a (radiation-reacted) 
	gravitational potential describing the scattering of equal-mass, non-spinning binary black holes.
	We urge the NR community to compute more sequences of scattering simulations, so as to extend this knowledge to a wider region of parameter space.
\end{abstract}

\maketitle

\section{Introduction}

The analysis of gravitational wave (GW) signals detected by the LIGO-Virgo-KAGRA (LVK) collaboration~\cite{LIGOScientific:2018mvr,LIGOScientific:2020ibl,LIGOScientific:2021djp,LIGOScientific:2014pky,VIRGO:2014yos,KAGRA:2020agh} relies on fast and accurate waveform models. The need
for accurate templates will be heightened when the third generation of interferometers~\cite{Reitze:2019iox,Punturo:2010zz,LISA:2017pwj} will start to detect GW events at higher rates and with improved signal-to-noise ratios.

Numerical Relativity (NR) simulations~\cite{Pretorius:2005gq, Campanelli:2005dd,Mroue:2013xna, Husa:2015iqa, Jani:2016wkt, Boyle:2019kee, Healy:2019jyf} play a key role in providing accurate solutions of Einstein's equations in the strong-field regime.
Their downside is the computational cost, which makes it impossible to use them directly in parameter estimation studies. NR surrogates~\cite{Field:2013cfa, Blackman:2017dfb, Varma:2019csw} partially
circumvent this problem, but they are limited in terms of parameter space coverage and waveform length.
All other GW approximants used by the LVK collaboration are, at least to some extent, based on analytical approximations. The most accurate ones are either models based on the effective-one-body (EOB) approach~\cite{Buonanno:1998gg,Buonanno:2000ef,Gamba:2021ydi,Ossokine:2020kjp} or phenomenological approximants~\cite{Khan:2019kot,Pratten:2020fqn,Garcia-Quiros:2020qpx,Pratten:2020ceb}. Both of these families are built using (approximate) analytical solutions of Einstein's equations and completed through the use of NR information. 

Historically, the prime analytical approximation method has been the post-Newtonian (PN) expansion~\cite{Blanchet:1989ki,Blanchet:2013haa,Damour:2014jta,Levi:2015uxa,Bini:2017wfr,Schafer:2018kuf,
Bini:2019nra,Bini:2020wpo,Bini:2020nsb,Bini:2020hmy,Antonelli:2020ybz}, which assumes small  
velocities ($v/c \ll 1$), together with weak fields  [$G M / (r c^2) \ll 1$]. A few years ago, the interest
in developing the post-Minkowskian approximation, which only assumes weak fields [$G M / (r c^2) \ll 1$],
and allows for arbitrary large velocities, has been pointed out ~\cite{Damour:2016gwp,Damour:2017zjx}. 
This triggered a lot of recent activity, using various approaches to gravitational scattering,
such as: scattering amplitudes (see, e.g.,~\cite{Cheung:2018wkq,Guevara:2018wpp,Kosower:2018adc,Bern:2019nnu,
Bern:2019crd,Bern:2019nnu,Bjerrum-Bohr:2019kec,
Bern:2021dqo,Bern:2021yeh,Bjerrum-Bohr:2021din,Manohar:2022dea,
Saketh:2021sri}); eikonalization (e.g.,~\cite{KoemansCollado:2019ggb,DiVecchia:2019kta,DiVecchia:2021bdo,DiVecchia:2022nna}); effective field theory (e.g.,~\cite{Kalin:2020mvi,Kalin:2020fhe,
Mougiakakos:2021ckm,Dlapa:2021npj,Dlapa:2021vgp,Kalin:2022hph,Dlapa:2022lmu}); and worldline (classical
or quantum) field theory (e.g.,~\cite{Mogull:2020sak,Riva:2021vnj,Jakobsen:2021smu,Jakobsen:2022psy}).

As the PM expansion is particularly suitable for describing scattering systems, 
it could be of help to improve GW models for eccentric and hyperbolic binaries signals~\cite{Chiaramello:2020ehz,Nagar:2021gss,Placidi:2021rkh,Khalil:2021txt,Ramos-Buades:2021adz,Nagar:2021xnh,Khalil:2022ylj}.
One expects that the LVK collaboration will eventually observe (and has possibly already observed~\cite{Romero-Shaw:2020thy,CalderonBustillo:2020odh,Gayathri:2020coq,Gamba:2021gap}) GW signals emitted by highly eccentric, capture or even hyperbolic binary systems.
Such observations (which are expected to be more common in third generation detectors) would improve our knowledge of BH formation and evolution~\cite{OLeary:2005vqo,OLeary:2008myb,Samsing:2013kua,Rodriguez:2016kxx,Belczynski:2016obo,Samsing:2017xmd}.

While PM results have been extended to take into account spin (e.g.,~\cite{Bini:2017xzy,Bini:2018ywr,Vines:2017hyw,Vines:2018gqi,Guevara:2019fsj,Kalin:2019inp,Kosmopoulos:2021zoq,Aoude:2022thd,Jakobsen:2022fcj,Bern:2020buy,Bern:2022kto,FebresCordero:2022jts}) and tidal (e.g.,~\cite{Bini:2020flp,Bern:2020uwk,Cheung:2020sdj,Kalin:2020lmz}) effects, in this paper we focus on non-spinning black hole (BH) binaries, for which 4PM-accurate results are available, both for conservative~\cite{Bern:2019nnu,Kalin:2020fhe,Bjerrum-Bohr:2021din,Bern:2021yeh,Dlapa:2021vgp} and radiation-reacted dynamics~\cite{Damour:2020tta,DiVecchia:2021ndb,Cho:2021arx,DiVecchia:2021bdo,Herrmann:2021tct,Bini:2021gat,Bini:2021qvf,Manohar:2022dea,Dlapa:2022lmu}.

The main aims of this paper are: (i) to compare\footnote{Our analysis has some overlap with Sec.~V of Ref.~\cite{Khalil:2022ylj}. However, the emphasis of our work is different and we use the information coming from the lowest impact-parameter (stronger-field) systems of Ref.~\cite{Damour:2014afa}, which were not taken into account in Ref.~\cite{Khalil:2022ylj}.} the equal-mass, nonspinning scattering simulations of Ref.~\cite{Damour:2014afa} to the analytical PM results of Refs.~\cite{Bern:2021yeh,Bern:2021dqo,Dlapa:2021vgp,Manohar:2022dea,Dlapa:2022lmu};
(ii) to propose a new resummation of the PM-expanded scattering angle incorporating the presence of a singularity at low impact parameters;
(iii) to use a new approach (introduced in Ref.~\cite{Damour:2017zjx}) for encoding the analytical PM information in the form of corresponding effective-one-body (EOB) radial gravitational potentials;
and (iv) to use Firsov's inversion formula~\cite{Landau:1960mec,Kalin:2019rwq} to extract a (radiation-reacted) gravitational potential directly from NR data.

\section{Post-Minkowskian scattering angle and effective-one-body}
\label{sec:scatt_angle}

Throughout this paper, we use units such that $G = c = 1$.
We consider a nonspinning BH binary system with masses $m_1, m_2$ and describe its dynamics with mass-rescaled coordinates and momenta:
\begin{align}
	r &\equiv R/M\,, \hspace{1cm} t \equiv T/M\,, \nonumber \\ 
	p_\alpha &\equiv P_\alpha/ \mu\,, \hspace{1cm} j \equiv J/(\mu M)\,,
\end{align}
where $M = m_1 + m_2$ is the total mass of the system and $\mu = (m_1 m_2)/M$ its reduced mass. We denote the symmetric mass ratio as $\nu \equiv \mu/M$.

\subsection{Post-Minkowskian-expanded scattering angle}

The PM-based study of the scattering of two body systems was initially done with the (implicit) assumption that the dynamics is conservative.
The discrepancy between the high-energy limit of the 3PM scattering angle first computed in Ref.~\cite{Bern:2019nnu} and the earlier result of Ref.~\cite{Amati:1990xe} attracted the attention on the need to take into account radiation reaction effects on gravitational scattering~\cite{Damour:2020tta,DiVecchia:2021ndb}.
In presence of radiation reaction, the scattering cannot be generally described only by a center-of-mass (c.m.) scattering angle but rather by the three coefficients of the decomposition of each ``impulse'' $\Delta p_1^\mu$, $\Delta p_2^\mu$ along a basis of three incoming four-vectors, e.g. $u^{-\mu}_{1}$, $u^{-\mu}_{2}$ and $\hat{b}_{12}^{\mu}$ ~\cite{Herrmann:2021tct,Bini:2021gat,Dlapa:2022lmu}.
However, if one can neglect the square of the recoil\footnote{As $\mathbf{P}^+$ is $O[G^3]$, the relative scattering angle used here, as well as its transcription in terms of a potential, is well-defined up to $O[G^6]$.}, $\mathbf{p_1^+} + \mathbf{p_2^+} = \mathbf{P^+} = - \mathbf{P}_{\rm rad}$, in the incoming c.m. frame ($\mathbf{p_1^-} + \mathbf{p_2^-} = \mathbf{P^-} = 0$),
one can define the \textit{relative} scattering angle $\chi_{\rm rel}$ (in the sense of both Ref.~\cite{Bini:2021gat} and Ref.~\cite{Dlapa:2022lmu}) as being the angle between the direction of the outgoing relative three-momentum $\mathbf{p^+}$ and the direction of the incoming relative three-momentum $\mathbf{p^-} = \mathbf{p_1^-} = -\mathbf{p_2^-}$.
Using Eq.~(3.29) in Ref.~\cite{Bini:2021gat}, $\mathbf{p^+}$ can be computed in terms of the outgoing individual four-momenta $(E_1^+,\mathbf{p}^+_1)$ and $(E_2^+,\mathbf{p}^+_2)$ as
\begin{equation}
\mathbf{p}^+ = \frac{E_2^+}{E_1^+ + E_2^+}\mathbf{p}_1^+ - \frac{E_1^+}{E_1^+ + E_2^+}\mathbf{p}_2^+ + O[(\mathbf{P}^+)^2]\, .
\end{equation}

In the following, the relative scattering angle $\chi_{\rm rel}$ will simply be denoted as $\chi$.
In the case considered here of an equal-mass system, the recoil $\mathbf{P}^+$ (proportional to $m_2-m_1$) vanishes and the c.m.-frame scattering angle is the common angle between $\mathbf{p}_1^+$ and $\mathbf{p}_1^-$, as well as between $\mathbf{p}_2^+$ and $\mathbf{p}_2^-$.

The (relative) scattering angle of nonspinning BH binaries can be PM-expanded as a power series in the inverse dimensionless angular momentum $j$ as
\begin{align}
	\label{eq:chiPM_1}
	\chi \left(\gamma, j\right) &= 2 \frac{\chi_1 \left(\gamma\right)}{j} + 2 \frac{\chi_2 \left(\gamma\right)}{j^2} + \nonumber \\
	&\quad + 2 \frac{\chi_3 \left(\gamma\right)}{j^3} + 2 \frac{\chi_4 \left(\gamma\right)}{j^4} + O\left[\frac{1}{j^5}\right],
\end{align}
where the expansion coefficients depend on the total energy, $E$, of the system in the incoming state. See Eq.~\eqref{eq:gamma} below for the relation between the energy $E$, the effective EOB energy $\mathcal{E}_{\rm eff}$, and the quantity $\gamma = -u_1^- \cdot u_2^-$, used in Eq.~\eqref{eq:chiPM_1}, which is the relative Lorentz factor of the two incoming worldlines.
The scattering angle at the $n$PM-order accuracy, $\chi_{n{\rm PM}}$, is then defined as
\begin{equation}
	\label{eq:chiPM}
	\chi_{n{\rm PM}}(\gamma,j) \equiv \sum_{i = 1}^{n} 2 \frac{\chi_i(\gamma)}{j^i}\,.
\end{equation}

The $\chi_i(\gamma)$ coefficients read
\begin{align}
		\label{eq:chi_i}
	\chi_1(\gamma) &= \frac{2 \gamma^2 - 1}{\sqrt{\gamma^2-1}}\,, \nonumber \\
	\chi_2(\gamma) &= \frac{3 \pi}{8} \frac{(5 \gamma^2 - 1)}{h(\gamma;\nu)}\,, \nonumber \\
	\chi_3(\gamma) &= \chi_3^{\rm cons}(\gamma) + \chi_3^{\rm rr}(\gamma)\,, \nonumber \\
	\chi_4(\gamma) &= \chi_4^{\rm cons}(\gamma) + \chi_4^{\rm rr, odd}(\gamma) + \chi_4^{\rm rr, even}(\gamma)\,.
\end{align}

Here we defined the rescaled energy $h(\gamma;\nu)$ as 
\begin{equation}
	\label{eq:h}
	h(\gamma;\nu) \equiv \frac{E}{M} = \sqrt{1 + 2 \nu (\gamma - 1)}\,,
\end{equation}
and, starting at 3PM, we separated the $\chi_i(\gamma)$ coefficients into conservative and radiation-reaction contributions.
The conservative 3PM contribution to the scattering angle reads~\cite{Bern:2019nnu,Kalin:2020fhe}
\begin{align}
	\chi_3^{\rm cons}(\gamma) &= \chi_3^{\rm Schw}(\gamma) - \frac{2 \nu \, p_\infty}{h^2(\gamma;\nu)} \bar{C}^{\rm cons}(\gamma)\,,
\end{align}
where 
\begin{equation}
	\chi_3^{\rm Schw}(\gamma) = \frac{64 p_\infty^6 + 72 p_\infty^4 + 12 p_\infty^2 - 1}{3 p_\infty^3}\,,
\end{equation}
with
\begin{equation}
	p_\infty \equiv \sqrt{\gamma^2 - 1}\,.
\end{equation}
The other building blocks entering $\chi_3^{\rm cons}(\gamma)$ are 
\begin{align}
	A(\gamma) &\equiv 2 \, {\rm arcsinh \sqrt{\frac{\gamma-1}{2}}},  \\
	\bar{C}^{\rm cons}(\gamma) &= \frac23 \gamma (14 \gamma^2 + 25) + 2(4 \gamma^4 - 12\gamma^2 - 3) \frac{A (\gamma)}{p_\infty}\,.
\end{align}

The radiation reaction contribution at 3PM order reads~\cite{Damour:2020tta,DiVecchia:2021ndb}
\begin{align}
\chi_3^{\rm rr}(\gamma) &= - \frac{2 \nu \, p_\infty}{h^2(\gamma;\nu)} \bar{C}^{\rm rad}(\gamma)\,,
\end{align}
where
\begin{align}
\bar{C}^{\rm rad}(\gamma) &= \frac{\gamma \, (2\gamma^2 - 1)^2}{3 (\gamma^2 - 1)^2} \left[\frac{5 \gamma^2 - 8}{\gamma} p_\infty + (9 - 6\gamma^2) A(\gamma)\right]\,.
\end{align}

As indicated in the last line of Eq.~\eqref{eq:chi_i}, the 4PM scattering angle is conveniently decomposed as the sum of three contributions: (i) the conservative contribution $\chi_{\rm 4}^{\rm cons}$~\cite{Bern:2021yeh,Dlapa:2021vgp}; (ii) the radiation-reaction contribution that is odd under time-reversal~\cite{Bini:2012ji,Bini:2021gat,Manohar:2022dea};
and (iii) the radiation-reaction contribution that is even under time-reversal~\cite{Dlapa:2022lmu,Bini:2022enm}. It has the structure
\begin{equation}
	\chi_4(\gamma;\nu) = \frac{1}{h^{3}(\gamma;\nu)} \left[\chi_4^{\rm Schw}(\gamma) + \nu\, \chi_4^{(1)}(\gamma) + \nu^2 \, \chi_4^{(2)}(\gamma)  \right]\,,
\end{equation}
where the test-mass contribution is
\begin{align}
	\chi_4^{\rm Schw}(\gamma) = \frac{105}{128}\pi\left(1-18\gamma^2+33\gamma^4\right) \, ,
\end{align}
and where the explicit forms of $\chi_4^{(1)}(\gamma)$ and $\chi_4^{(2)}(\gamma)$ can be found in the ancillary file of Ref.~\cite{Dlapa:2022lmu}.

\subsection{Effective-one-body mass-shell condition and potential}

The EOB formalism~\cite{Buonanno:1998gg,Buonanno:2000ef,Damour:2000we} was initially conceived as a way of mapping the conservative two body dynamics onto a one-body Hamiltonian describing the relative motion of the two objects (in the c.m. frame). The ``real'' c.m. Hamiltonian is related to the ``effective'' Hamiltonian through
\begin{equation}
	H_{\rm real} = M \sqrt{1+2\nu\left(\frac{H_{\rm eff}}{\mu}-1\right)}\, .
\end{equation}

The comparison to Eq.~\eqref{eq:h} shows that the Lorentz-factor variable $\gamma = -u^-_1 \cdot u^-_2$ is equal to the dimensionless effective energy of the system $\hat{\mathcal{E}}_{\rm eff} \equiv \mathcal{E}_{\rm eff}/\mu$, i.e.
\begin{equation}
	\label{eq:gamma}
	\gamma = \hat{\mathcal{E}}_{\rm eff} = \frac{\left(E_{\rm real}\right)^2-m_1^2-m_2^2}{2 m_1 m_2}\, .
\end{equation}

The EOB dynamics is encapsulated in a mass-shell condition of the general form
\begin{equation}
	\label{eq:massshell}
	\mu^2 + g_{\rm eff}^{\mu \nu} P_\mu P_\nu + Q(X^\mu,P_\mu) = 0\, ,
\end{equation} 
where $Q(X^\mu,P_\mu)$ is a Finsler-type term accounting for higher-than-quadratic momenta contributions. In rescaled variables, $p_\alpha \equiv P_\alpha/ \mu$, $x^\alpha \equiv X^\alpha/M$, and $\hat{Q} \equiv Q/\mu^2$, this becomes
\begin{equation}
		\label{eq:mass-shell}
	1 + g_{\rm eff}^{\mu \nu} p_\mu p_\nu + \hat{Q}(x^\mu,p_\mu) = 0\, .
\end{equation}
The quadratic-in-momenta term, $g_{\rm eff}^{\mu \nu} p_\mu p_\nu$, defines an effective 
metric of the general form
\begin{equation}
	\label{eq:metric}
	g_{\mu \nu}^{\rm eff} dx^\mu dx^\nu = -A(r) dt^2 + B(r)dr^2 + C(r)d\Omega^2\, .
\end{equation}  
The symmetries of Eq.~\eqref{eq:mass-shell} imply the existence of two constants of motion, energy and angular momentum, namely $\hat{\mathcal{E}}_{\rm eff} = -p_0$ and $j = p_\varphi$.

The general mass-shell condition, Eq.~\eqref{eq:mass-shell}, has been used in the EOB literature in various gauges. 
Following the post-Schwarzschild approach of Ref.~\cite{Damour:2017zjx}, it is convenient to fix the metric $g_{\mu \nu}^{\rm eff}$ to be the Schwarzschild metric.
Writing the latter metric in terms of isotropic coordinates [with a radial coordinate\footnote{The isotropic radial coordinate $\bar{r}$ is related to the usual Schwarzschild-like $r$ by $r = \bar{r}\left[1+1/(2 \bar{r})\right]^2$.} $\bar{r}$ that satisfies $\bar{C}(\bar{r}) = \bar{r}^2 \bar{B}(\bar{r})$]~\cite{Damour:2017zjx,Damour:2019lcq} and expressing the momentum dependence of $\hat{Q}(x^\mu,p_\mu)$ only in terms of $p_0 = - \gamma$,
leads to a mass-shell condition of the form
\begin{equation}
	\label{eq:p2}
	p_{\bar{r}}^2 + \frac{j^2}{\bar{r}^2} =p_\infty^2 + w(\bar{r},\gamma) \, .
\end{equation}

This mass-shell condition has the useful property of reducing the two-body dynamics to the Newtonian dynamics of a non-relativistic particle moving in the radial potential $-w(\bar{r},\gamma)$.
When considering motions with a given $j$, the mass-shell condition \eqref{eq:p2} can be re-written as
\begin{equation}
	p_{\bar{r}}^2 = p_\infty^2 - V(\bar{r},\gamma,j)\, ,
\end{equation}
where we introduced the effective potential 
\begin{equation}
	\label{eq:V}
	V(\bar{r},\gamma,j) = \frac{j^2}{\bar{r}^2} - w(\bar{r},\gamma)\, .
\end{equation}
This effective potential features a usual-looking centrifugal potential $j^2/\bar{r}^2$ [containing the entire $j$ dependence of $V(\bar{r},\gamma,j)$] together with the original (energy-dependent) radial potential $-w(\bar{r},\gamma)$.

Recalling that $p_\infty^2 = \gamma^2 -1$, the radial potential $w(\bar{r},\gamma)$ is obtained, combining Eqs.~\eqref{eq:mass-shell}-\eqref{eq:metric}, as
\begin{align}
	\label{eq:w}
	w(\bar{r}, \gamma) &= \gamma^2 \left[\frac{{\bar B}(\bar{r})}{{\bar A}(\bar{r})}-1 \right] + \nonumber \\
	&\quad - \left[{\bar B}(\bar{r})-1 \right] - {\bar B}(\bar{r}) \hat{Q}(\bar{r}, \gamma)\,.
\end{align}

In this formula, 
\begin{equation}
	\label{eq:ASchw}
	{\bar A}(\bar{r}) = \left(\frac{1- \frac{1}{2\bar{r}}}{1+ \frac{1}{2\bar{r}}}\right)^2\,, \hspace{0.75cm}
	{\bar B}(\bar{r}) =\left(1+ \frac{1}{2\bar{r}} \right)^4\,,
\end{equation}
describe the Schwarzschild metric in isotropic coordinates. Instead, $\hat{Q}(\bar{r},\gamma)$ admits a PM expansion 
in the inverse of the radius $\frac{1}{\bar{r}}$ ($= \frac{GM}{\bar{R}}$ in unrescaled units) of the form
\begin{equation}
	\label{eq:Q}
\hat{Q}(\bar{r}, \gamma) = \frac{\bar{q}_2(\gamma)}{\bar{r}^2} + \frac{\bar{q}_3(\gamma)}{\bar{r}^3} + \frac{\bar{q}_4(\gamma)}{\bar{r}^4} + O\left[\frac{1}{\bar{r}^5} \right]\,.
\end{equation}
The first term in this expansion is at at 2PM order, i.e. $\propto \frac{G^2 M^2}{\bar{R}^2}$. 

Such an isotropic-coordinate description of the EOB dynamics with an energy-dependent potential was introduced in Ref.~\cite{Damour:2017zjx}. This EOB potential is similar to the isotropic-gauge
EFT-type potential later-introduced in Ref.~\cite{Cheung:2018wkq}, and used e.g. in Ref.~\cite{Kalin:2019rwq}. The relation between these two types of potential is discussed in Appendix~A of Ref.~\cite{Damour:2019lcq} (see also Ref.~\cite{Antonelli:2019ytb}).

Inserting the expansion~\eqref{eq:Q} into Eq.~\eqref{eq:w}, and re-expanding\footnote{One could also define a potential where the Schwarzschild metric coefficients $\bar{A}$ and $\bar{B}$ are \textit{not} expanded in powers of $\bar{u}$.} in powers of $\bar{u}$, defines the PM-expansion of the (energy-dependent) radial potential in the form
\begin{equation}
	\label{eq:wr}
w(\bar{r}, \gamma) = \frac{w_1(\gamma)}{\bar{r}} + \frac{w_2(\gamma)}{\bar{r}^2} +\frac{w_3(\gamma)}{\bar{r}^3} +\frac{w_4(\gamma)}{\bar{r}^4} + O\left[\frac{1}{\bar{r}^5}\right]\,.
\end{equation}

This energy-dependent potential encodes the \textit{attractive} gravitational interaction between the two bodies. It is a relativistic generalization of the Newtonian potential $U = +(G M)/R$.
In the following, we will introduce the sequence of $n$PM potentials, defined as 
\begin{equation}
\label{eq:wPM}
w_{n{\rm PM}}(\bar{r},\gamma) \equiv \sum_{i = 1}^{n} \frac{w_i(\gamma)}{ \bar{r}^i}\,.
\end{equation}

We can then compute the scattering angle $\chi \left(\gamma, j\right)$ as
\begin{equation}
\label{eq:chi_pr}
\pi + \chi \left(\gamma, j\right) = - \int_{-\infty}^{+\infty} d\bar{r}\, \frac{\partial p_{\bar{r}}\left(\bar{r},j,\gamma\right)}{\partial j}\,,
\end{equation}
where the limits of integration $\bar{r} = \mp \infty$ are to be interpreted as the incoming and final states (at $t = \mp \infty$ respectively).

In isotropic coordinates, it is easy to invert the mass-shell condition to obtain the radial momentum as 
\begin{align}
p_{\bar{r}}\left(\bar{r},j,\gamma\right) 
 &= \pm \sqrt{p_\infty^2 - V(\bar{r},\gamma,j)}\,, \nonumber \\
 &= \pm \sqrt{p_\infty^2 + w(\bar{r},\gamma) - \frac{j^2}{\bar{r}^2}}\,,
\end{align}
where the square root has a negative sign along the ingoing trajectory and a positive one during the outgoing motion. Its derivative with respect to $j$ reads
\begin{equation}
\frac{\partial p_{\bar{r}}\left(\bar{r},j,\gamma\right)}{\partial j} = - \frac{j}{\bar{r}^2 \, p_{\bar{r}}\left(\bar{r},j,\gamma\right)}\, .
\end{equation}

Eq.~\eqref{eq:chi_pr} thus becomes
\begin{equation}
\label{eq:chi_w}
\pi + \chi \left(\gamma, j\right) = 2\, j \int_{\bar{r}_{\rm min}(\gamma,j)}^{+\infty} \frac{d\bar{r}}{\bar{r}^2}\, \frac{1}{\sqrt{p_\infty^2 + w(\bar{r},\gamma) - j^2/\bar{r}^2}}\, , 
\end{equation}
in which we introduced the radial turning point $\bar{r}_{\rm min}(\gamma,j)$, defined, in the present scattering context, 
as the largest root of the radial momentum $p_{\bar{r}}\left(\bar{r},j,\gamma\right)$.

In the following, it will often be convenient to replace $\bar{r}$ by its inverse
\begin{equation}
	\bar{u} \equiv \frac{1}{\bar{r}}\, ,
\end{equation}
so that Eq.~\eqref{eq:chi_w} reads
\begin{equation}
	\label{eq:chi_wu}
\pi + \chi \left(\gamma, j\right) = 2\, j \int_{0}^{\bar{u}_{\rm max}(\gamma,j)} \, \frac{d\bar{u}}{\sqrt{p_\infty^2 + w(\bar{u},\gamma) - j^2 \bar{u}^2}}\, ,
\end{equation}
where $\bar{u}_{\rm max}(\gamma,j) \equiv 1/\bar{r}_{\rm min}(\gamma,j)$.

Inserting the PM-expanded $w(\bar{u}, \gamma)$, Eq.~\eqref{eq:wr}, into Eq.~\eqref{eq:chi_wu}, PM-expanding the integrand, and taking the \textit{partie finie} of the resulting (divergent) integrals yields the relation\footnote{An alternative route to connect the $\chi_i(\gamma)$ to the $w_i(\gamma)$ is to go through the coefficients $\bar{q}_i(\gamma)$ of Eq.~\eqref{eq:Q}. See Ref.~\cite{Damour:2017zjx}.} between the PM coefficients of the scattering angle, Eq.~\eqref{eq:chiPM}, to the ones of the radial potential~\cite{Damour:1988mr,Damour:2019lcq}. Namely,
\begin{align}
\label{eq:coeffs_wchi}
\chi_1(\gamma) &= \frac12 \frac{w_1(\gamma)}{p_\infty}\,, \nonumber \\
\chi_2(\gamma) &= \frac \pi 4 w_2(\gamma)\,, \nonumber \\
\chi_3(\gamma) &= - \frac{1}{24} \left[\frac{w_1(\gamma)}{p_\infty}\right]^3 + \frac12 \frac{w_1(\gamma) w_2(\gamma)}{p_\infty} + p_\infty w_3(\gamma)\,, \nonumber \\
\chi_4(\gamma) &= \frac{3 \pi}{8} \left[\frac12 w_2^2(\gamma) + w_1(\gamma) w_3(\gamma) + p_\infty^2 w_4(\gamma)\right]\,.
\end{align}

The latter relations between the scattering angle and corresponding potential are initially defined by considering the conservative dynamics of binary systems. 
Following Ref.~\cite{Damour:2017zjx}, where the radiation-reacted high-energy dynamics of Ref.~\cite{Amati:1990xe} was transcribed in terms of an equivalent effective metric (see Sec. VII there), we similarly extend here the use of the relations \eqref{eq:coeffs_wchi} to transcribe the radiation-reacted 4PM scattering angle, Eq.~\eqref{eq:chiPM_1}, in terms of an equivalent effective potential.
The latter potential will be a function of the \textit{incoming} energy and angular momentum of the binary system,
but will contain additional radiation-reacted contributions that effectively take into account the variation
of the energy and angular momentum of the system during the scattering (see e.g. Sec. III of Ref.~\cite{Bini:2021gat}).
As we are going to see, while the conservative contributions to the latter potential are time-even (even in velocities), the fact that the potential also transcribes radiative effects will be reflected in the presence of time-odd terms.

Inserting the explicit values of $\chi_i(\gamma)$, Eq.~\eqref{eq:chi_i}, and solving for the $w_i(\gamma)$'s then yields
\begin{align}
	\label{eq:chi_to_w}
	w_1(\gamma) &= 2(2 \gamma^2 - 1)\,, \nonumber \\
	w_2(\gamma) &= \frac{3}{2} \frac{(5 \gamma^2 - 1)}{h(\gamma;\nu)}\,, \nonumber \\
	w_3(\gamma) &= w_3^{\rm cons}(\gamma) + w_3^{\rm rr}(\gamma)\,, \nonumber \\
	w_4(\gamma) &= w_4^{\rm cons}(\gamma) + w_4^{\rm rr, odd}(\gamma) + w_4^{\rm rr, even}(\gamma)\,.
\end{align}

Here we have separated the 3PM coefficient $w_3(\gamma)$ in a part $w_3^{\rm cons}(\gamma)$, computed from $\chi_1(\gamma)$, $\chi_2(\gamma)$ and $\chi_3^{\rm cons}(\gamma)$, namely
\begin{align}
	w_3^{\rm cons}(\gamma) &= 9 \gamma^2 -\frac12 - B(\gamma)\left[\frac{1}{h(\gamma;\nu)} - 1\right] \nonumber \\
	                                         &- \frac{2 \nu}{h^2(\gamma;\nu)} \bar{C}^{\rm cons}(\gamma)\, ,
\end{align}
where 
\begin{align}
B(\gamma) &= \frac32 \frac{(2 \gamma^2 - 1)(5 \gamma^2 - 1)}{\gamma^2 - 1}\, ,
\end{align}
and a part $w_3^{\rm rr}(\gamma)$ proportional to the radiative part of the scattering angle, $\chi_3^{\rm rr}(\gamma)$, namely
\begin{align}
w_3^{\rm rr}(\gamma) &= - \frac{2 \nu}{h^2(\gamma;\nu)} \bar{C}^{\rm rad}(\gamma)\, .
\end{align}
The radiative origin of $w_3^{\rm rr}(\gamma)$ manifests itself in the fact that, contrary to $w_3^{\rm cons}(\gamma)$, $w_3^{\rm rr}(\gamma)$ is odd under time (and velocity) reversal.

The 4PM coefficient $w_4(\gamma)$ is obtained by inverting Eq.~\eqref{eq:coeffs_wchi}. 
Its explicit explicit expression is intricate, and consists of three contributions: a conservative (time-even) piece, $w_4^{\rm cons}(\gamma)$; a time-odd radiation-reaction piece, 
$w_4^{\rm rr, odd}(\gamma)$ [which is a linear combination of $\chi_3^{\rm rr}(\gamma)$ and $\chi_4^{\rm rr, odd}(\gamma)$]; and a time-even radiation-reaction piece, $w_4^{\rm rr, even}(\gamma)$ [which is proportional to $\chi_4^{\rm rr, even}(\gamma)$].
The latter contribution $w_4^{\rm rr, even}(\gamma)$ is proportional to second-order radiation-reaction effects (in the sense of Sec. X of Ref.~\cite{Bini:2021gat}, of Eq.~(16) in the supplemental material of Ref.~\cite{Dlapa:2022lmu}, and of Eq.~(12.35) in Ref.~\cite{Bini:2022enm}).

\section{Defining a sequence of scattering angles associated to the effective-one-body radial potentials}
\label{sec:chiw}

\subsection{On the angular-momentum dependence of the $V$ potentials}

\begin{figure}[t]
	\includegraphics[width=0.48\textwidth]{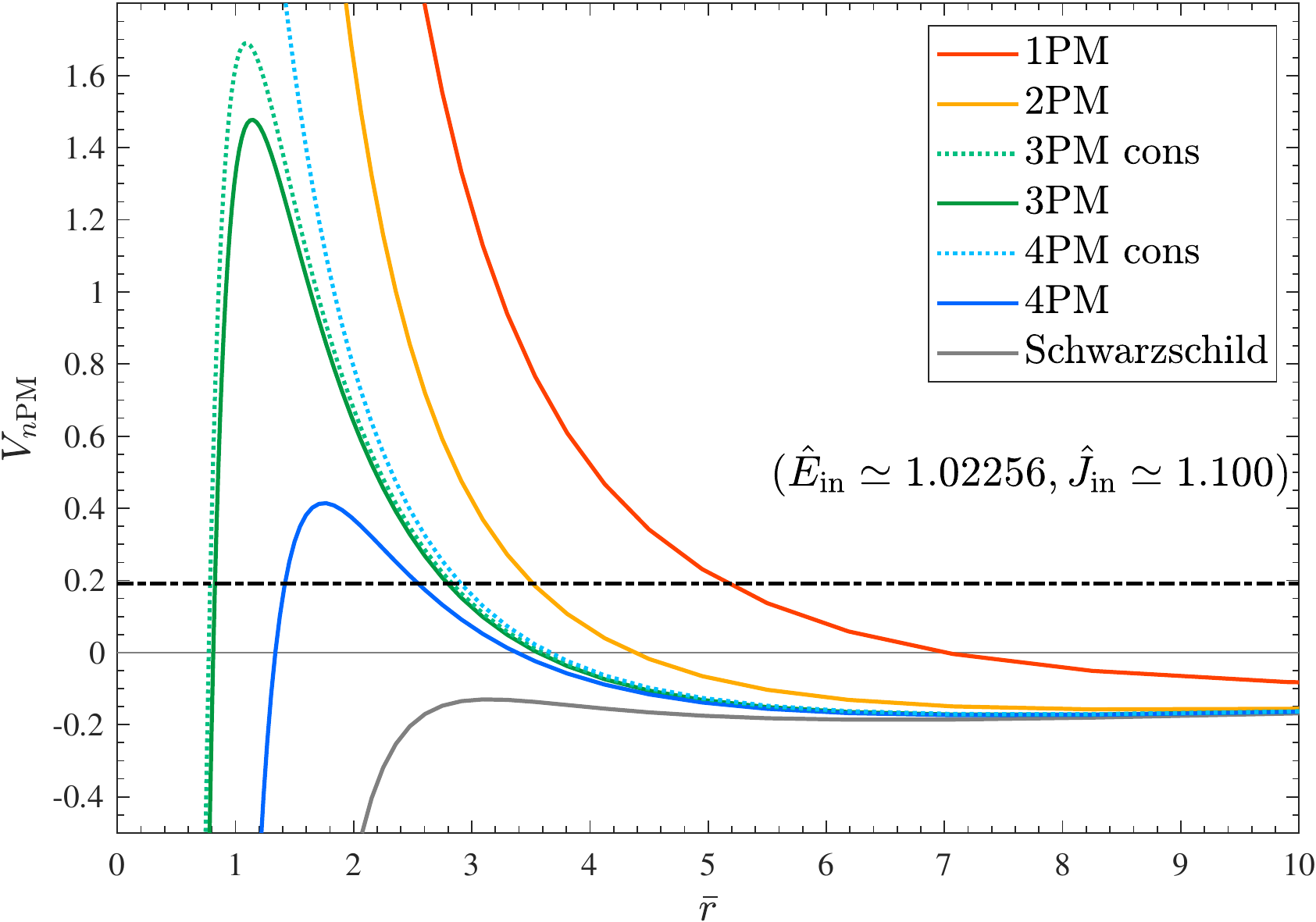}
	\caption{
		\label{fig:wPM2}
		PM gravitational potential $V_{n{\rm PM}}$ at different perturbative orders.
		Energy and angular momentum are fixed to the first simulation of Ref.~\cite{Damour:2014afa}, i.e. $\hat{E}_{\rm in} \equiv E_{\rm in}/M \simeq 1.02256$ and $\hat{J}_{\rm in} \equiv J_{\rm in}/M^2 \simeq 1.100$.		
		The black, dot-dashed horizontal line marks the corresponding value of $p_\infty^2$.
		The high-order PM corrections tend to make the (radiation-reacted) radial potential more and more attractive.
		The system scatters in every PM potential, while it would plunge in the Schwarzschild case (plotted as a reference).
	}
\end{figure}

PM expansions are well adapted to describing the large-$j$ (weak-field) behavior of the scattering angle $\chi$.
However, we expect $\chi(\gamma,j)$ to have some type of singularity below a critical value of $j$.
In our present framework, the scattering at the $n$PM order is defined by an effective energy-dependent potential (which can include radiation-reaction effects),
of the form 
\begin{equation}
	V_{n{\rm PM}}(\bar{r},\gamma,j) \equiv \frac{j^2}{\bar{r}^2} - w_{n{\rm PM}}(\bar{r},\gamma)\, .
\end{equation}
The fact that the centrifugal barrier term $j^2/\bar{r}^2$ is separate from the attractive potential $- w_{n{\rm PM}}(\bar{r},\gamma)$ means that, depending on the relative magnitudes of the radial potential and of the centrifugal barrier, the system will either scatter (finite $\chi$; possibly with zoom-whirl behavior) or plunge ($\chi$ undefined).
This is illustrated in Fig.~\ref{fig:wPM2} for specific values of $\gamma$ and $j$, corresponding to the smallest impact-parameter NR simulation of Ref.~\cite{Damour:2014afa}, that we will use below.

The bottom curve in Fig.~\ref{fig:wPM2} displays the potential $V^{\rm Schw}(\bar{r},\gamma,j)$ corresponding to the exact (un-expanded) Schwarzschild potential $w^{\rm Schw}(\gamma,\bar{r})$ (test-mass limit) defined by Eqs.~\eqref{eq:w}-\eqref{eq:ASchw}.
In this case, the horizontal dash-dotted curve (corresponding to $p_\infty^2$) is above $V^{\rm Schw}$, which means that the centrifugal contribution is not strong enough to allow the system to scatter (contrary to the result of the NR simulation), but would lead to a plunge.
By contrast, all the other PM-expanded potentials lead to scattering. This is due to the fact that the $-w_{n{\rm PM}}(\gamma,\bar{r})$ radial potentials are \textit{less attractive} than $-w^{\rm Schw}(\gamma,\bar{r})$.

The first two PM potentials are somewhat exceptional. 
In the 1PM potential,
\begin{equation}
V_{\rm 1PM}(\bar{r},\gamma,j) = \frac{j^2}{\bar{r}^2} - \frac{2(2 \gamma^2 - 1)}{\bar{r}}\, ,
\end{equation}
 the centrifugal barrier ultimately dominates over the attractive $1/\bar{r}$ potential for any $j\neq0$.
Instead, in the 2PM potential
\begin{align}
	V_{\rm 2PM}(\bar{r},\gamma,j) &= \left[j^2 - w_2(\gamma)\right]\frac{1}{\bar{r}^2} - \frac{w_1(\gamma)}{\bar{r}}\,, \nonumber \\
	&= \left[j^2 - \frac{3}{2} \frac{(5 \gamma^2 - 1)}{h(\gamma;\nu)}\right]\frac{1}{\bar{r}^2} - \frac{2(2 \gamma^2 - 1)}{\bar{r}}\, , 
\end{align}
the small-$\bar{r}$ behavior is always repulsive (leading to scatter) if $j^2 > w_2(\gamma)$, 
and always attractive (leading to plunge) if $j^2 < w_2(\gamma)$.

For higher PM orders (except for the conservative 4PM case discussed below), the radial potential $-w_{n{\rm PM}}(\bar{r},\gamma)$ is always attractive and wins over the centrifugal potential $j^2/\bar{r}^2$ at small $\bar{r}$. Then, there exists an energy-dependent \textit{critical angular momentum}, say $j_0(\gamma,\nu)$, such that the system scatters for $j > j_0$ and plunges for $j < j_0$.
When $j \rightarrow j_{0}^{+}$, the scattering angle $\chi(\gamma,j)$ has a logarithmic singularity, as discussed in the next sections. 

The conservative 4PM radial potential $-w_{4{\rm PM cons}}(\bar{r},\gamma)$ is exceptional in that the coefficient of $1/\bar{r}^4$, i.e. $w_4^{\rm cons}(\gamma;\nu)$, is not always positive. 
More precisely, $w_4^{\rm cons}(\gamma;\nu)$ is positive if $\nu \lesssim 0.2$ and $\gamma \gtrsim 1.425$, but becomes negative in a sub-region of the rectangle $0.2\lesssim\nu\le0.25,1\leq\gamma\lesssim1.425$.
Our NR data values ($\nu=1/4$, $\gamma \simeq 1.09136$) happen to fall in the region when $w_4^{\rm cons}$ is negative.
When this happens, $-w_{4{\rm PM cons}}(\bar{r},\gamma)$ features an (unphysical) repulsive core at small distances.
For our NR parameters, there still exists a critical angular momentum ($j_0 \approx 3.9983$) leading to a logarithmic singularity in the scattering angle.
However, when $j < j_0$, the system does not plunge but instead scatters against the repulsive core, so that the scattering angle $\chi$ monotonically decreases with $j$ (even becoming negative) and tends to $-\pi$ in the head-on limit, $j\rightarrow 0$.

\subsection{Singularity of $\chi(\gamma,j)$ in the test-mass limit}

As a warm-up, let us briefly discuss the singularity structure of $\chi(\gamma,j)$ in the test-mass limit.
In this case, it is easier to evaluate $\chi(\gamma,j)$ by using a Schwarzschild radial coordinate $r$, rather than the isotropic $\bar{r}$.
This leads to a scattering angle given by
\begin{align}
	\label{eq:chi_Schw}
	\chi^{\rm Schw} \left(\gamma, j\right) &= -\pi + \frac{2 \sqrt{2}}{\sqrt{u_3 - u_1}} \times \nonumber \\
	&\quad \times \left[K(k_{\rm Schw}^2) - F\left(\sin \varphi_{\rm Schw}; k_{\rm Schw}^2\right)\right]\,,
\end{align}
where $K$ and $F$ are the complete and incomplete elliptic integrals of the fist kind, defined as
\begin{align}
	F(\sin \varphi; k^2) &\equiv \int_{0}^{\varphi} \frac{d \theta}{\sqrt{1 - k^2 \sin^2 \theta}}\,, \nonumber \\
	K(k^2) &\equiv F(1; k^2)\,.
\end{align}

The elliptic parameters entering Eq.~\eqref{eq:chi_Schw} are
\begin{align}
	k_{\rm Schw}^2 &= \frac{u_2-u_1}{u_3-u_1}\,, \nonumber \\
	\sin \varphi_{\rm Schw} &= \sqrt{\frac{- u_1}{u_2 - u_1}}\,,
\end{align}
where $u_i = 1/r_i$ are the three roots of $p_r^2 = 0$ in Schwarzschild coordinates.
In our scattering situation, the three solutions satisfy $u_1 < 0 < u_2 < u_3$.

The critical angular momentum $j_0^{\rm Schw}$ corresponds to the case where $u_2 = u_3$, and is given by 
  \begin{equation}
	j_0^{\rm Schw} = \sqrt{\frac{1}{u_0^2} \left(\frac{\gamma^2}{1-2 u_0}-1\right)}\,, 
\end{equation}
where 
\begin{equation}
	u_0 = \frac{4-3 \gamma^2 + \gamma \sqrt{9 \gamma^2 - 8}}{8}\,.
\end{equation}

In the limit $j \rightarrow j_0^{\rm Schw+}$, $k_{\rm Schw}^2 \rightarrow 1^{-}$ and the complete elliptic integral $K(k_{\rm Schw}^2)$ diverges logarithmically, so that 
\begin{equation}
	\chi^{\rm Schw} \left(\gamma, j\right) \overset{\hspace{0.2cm} j \rightarrow j_0^{+}}{\approx} \frac{2}{\left(1-\frac{12}{j^2}\right)^{\frac{1}{4}}}\, \ln\left[\frac{1}{1-\frac{j_0(\gamma)}{j}}\right]\,.
\end{equation}

\subsection{Map between $\chi(\gamma,j)$ and the effective-one-body PM radial potentials $w_{n{\rm PM}}$}

Let us now compute the explicit expression of the scattering angle corresponding to a given $n$PM radial potential $w_{n{\rm PM}}(\bar{r},\gamma)$, as defined in Eq.~\eqref{eq:wPM}.
Explicitly we define 
\begin{equation}
	\label{eq:chi_wPM}
	\chi^{w\,{\rm eob}}_{n{\rm PM}} \left(\gamma, j\right) \equiv 2\, j \int_{0}^{\bar{u}_{\rm max}(\gamma,j)} \hspace{-0.75cm} \frac{d\bar{u}}{\sqrt{p_\infty^2 + w_{n{\rm PM}}(\bar{u},\gamma) - j^2 \bar{u}^2}} -\pi \,.
\end{equation}

On the right-hand side of this definition, enters the $n$PM-expanded radial potential $w_{n{\rm PM}}(\bar{r},\gamma) \sim 1/\bar{r} + \cdots + 1/\bar{r}^n$.
The corresponding nonlinear transformation, 
\begin{equation}
	w_{n{\rm PM}}(\bar{r},\gamma) \longrightarrow \chi^{w\,{\rm eob}}_{n{\rm PM}} \left(\gamma, j\right)\,,
\end{equation}
defines a new sequence of scattering angles, such that the $n^{\rm th}$ angle, $\chi^{w\,{\rm eob}}_{n{\rm PM}} \left(\gamma, j\right)$, incorporates analytical PM information up to the $n$PM order included.
A similar sequence of scattering angles incorporating analytical information up to some given PM order has been introduced in Sec. 4.2 of Ref.~\cite{Kalin:2019rwq}. The latter reference used a framework different from our EOB one, namely the PM-expansion of the two-body potential introduced in Ref.~\cite{Cheung:2018wkq}.

Each $\chi^{w\,{\rm eob}}_{n{\rm PM}}\left(\gamma, j\right)$ defines a function of $j$ which differs from the corresponding PM-expanded $\chi_{n{\rm PM}}\left(\gamma, j\right) \sim 1/j + \cdots + 1/j^n$.
It is a fully nonlinear function of $j$, whose first $n$ terms of its large-$j$ expansion coincide with $\chi_{n{\rm PM}}$.
The sequence $\chi^{w\,{\rm eob}}_{n{\rm PM}}$ defines a \textit{resummation} ($w^{\rm eob}$-resummation) of the PM-expanded $\chi_{n{\rm PM}}$.

In the following, we study the properties of the $w^{\rm eob}$-resummed sequence $\chi^{w\,{\rm eob}}_{n{\rm PM}}$ and show how it improves the agreement with numerical data, notably because it incorporates a singular behavior of $\chi(j)$ at some critical $j_0^{w_{n{\rm PM}}}$. 

Let us explicate the values of $\chi^{w\,{\rm eob}}_{n{\rm PM}}$ up to $n=4$.
At 1PM and 2PM, the radicand entering the denominator of the integral is quadratic in $\bar{u}$ and the integration is trivial, yielding
\begin{align}
\chi^{w\,{\rm eob}}_{1{\rm PM}} \left(\gamma, j\right) &= - \pi \nonumber \\
&\hspace{-1.5cm}+4 \arctan \left[\sqrt{\frac{\sqrt{w_1^2(\gamma) + 4\, p_\infty^2\, j^2} + w_1(\gamma)}{\sqrt{w_1^2(\gamma) + 4\, p_\infty^2\, j^2} - w_1(\gamma)}}\right]\,, \nonumber \\
	\chi^{w\,{\rm eob}}_{2{\rm PM}} \left(\gamma, j\right) &= -\pi + \frac{4\, j}{\sqrt{j^2 - w_2(\gamma)}} \times \nonumber \\
	&\hspace{-1.5cm}\times \arctan \left\{\sqrt{\frac{\sqrt{w_1^2(\gamma) - 4\, p_\infty^2 \left[w_2(\gamma)-j^2\right]} + w_1(\gamma)}{\sqrt{w_1^2(\gamma) - 4\, p_\infty^2 \left[w_2(\gamma)-j^2\right]} - w_1(\gamma)}}\right\}\,.
\end{align}
As already said, $\chi^{w\,{\rm eob}}_{1{\rm PM}}$ is defined for any $j$ and has no  singularity.
Instead, $\chi^{w\,{\rm eob}}_{2{\rm PM}}$ is only defined for angular momenta $j^2 \geq w_2(\gamma)$, below which the system plunges instead of scattering. 
In the limit $j^2 \rightarrow w_2(\gamma)^{+}$, $\chi^{w\,{\rm eob}}_{2{\rm PM}}$ has a power-law singularity $\chi^{w\,{\rm eob}}_{2{\rm PM}} \approx \frac{2 \pi j}{\sqrt{j^2-w_2(\gamma)}}$.

At 3PM and 4PM, the scattering angle can be expressed as a combination of elliptic integrals (similarly to the Schwarzschild case).
At 3PM, for high-enough angular momenta, $p_{\bar{r}}^2$ has (generally) three real roots $(\bar{u}_1,\bar{u}_2,\bar{u}_3)$, with $\bar{u}_1 < 0 < \bar{u}_2 \leq \bar{u}_3$ (so that $\bar{u}_{\rm max} \equiv \bar{u}_2$), and $\chi^{w\,{\rm eob}}_{3{\rm PM}}$ reads
\begin{align}
	\chi^{w\,{\rm eob}}_{3{\rm PM}} \left(\gamma, j\right) &= -\pi + \frac{4\, j}{\sqrt{w_3(\gamma)(\bar{u}_3 - \bar{u}_1)}} \times \nonumber \\
	&\quad \times \left[K(k_{\rm 3PM}^2) - F\left(\sin \varphi_{\rm 3PM}; k_{\rm 3PM}^2\right)\right]\,,
\end{align}
where 
\begin{align}
	k_{\rm 3PM}^2 &= \frac{\bar{u}_2-\bar{u}_1}{\bar{u}_3-\bar{u}_1}\,, \nonumber \\
	\sin \varphi_{\rm 3PM} &= \sqrt{\frac{- \bar{u}_1}{\bar{u}_2 - \bar{u}_1}}\,.
\end{align}

This formula is valid only when $k_{\rm 3PM}^2 < 1$ (i.e. $\bar{u}_3 > \bar{u}_2$). Like in the test-mass limit, $\chi^{w\,{\rm eob}}_{3{\rm PM}}$ diverges logarithmically in the limit $\bar{u}_3 \rightarrow \bar{u}_2$, determining the smallest angular momentum for which scattering occurs. 

If the 4PM radial potential $w_{4{\rm PM}}(\bar{u},\gamma)$ is such that $p_{\bar{r}}^2$ has 4 real roots $(\bar{u}_1,\bar{u}_2,\bar{u}_3,\bar{u}_4)$, with $\bar{u}_1 < \bar{u}_2 < 0 < \bar{u}_3 \leq \bar{u}_4$ (and $\bar{u}_{\rm max} \equiv \bar{u}_3$), we obtain\footnote{This is the case for $w_4(\gamma)$ but not for $w_4^{\rm cons}(\gamma)$ for the considered energy and mass ratio. In this case, there is only one negative real root and the result can be written as a different combination of elliptic integrals.}
\begin{align}
	\chi^{w\,{\rm eob}}_{4{\rm PM}} \left(\gamma, j\right) &= -\pi + \frac{4\, j}{\sqrt{w_4(\gamma)(\bar{u}_3 - \bar{u}_1)(\bar{u}_4 - \bar{u}_2)}} \times \nonumber \\
	&\quad \times \left[K(k_{\rm 4PM}^2) - F\left(\sin \varphi_{\rm 4PM}; k_{\rm 4PM}^2\right)\right]\,,
\end{align}
with
\begin{align}
	k_{\rm 4PM}^2 &= \frac{(\bar{u}_4-\bar{u}_1)(\bar{u}_3-\bar{u}_2)}{(\bar{u}_4-\bar{u}_2)(\bar{u}_3-\bar{u}_1)}\,, \nonumber \\
	\sin \varphi_{\rm 4PM} &= \sqrt{\frac{\bar{u}_2(\bar{u}_3-\bar{u}_1)}{\bar{u}_1(\bar{u}_3-\bar{u}_2)}}.
\end{align}

This formula is again valid only for $\bar{u}_3 < \bar{u}_4$.
The critical $j_0$ is determined by the coalescence $\bar{u}_3 \rightarrow \bar{u}_4$\,, which yields $k_{\rm 4PM}^2 \rightarrow 1$.

At higher PM orders, the computations will be similar but will involve hyper-elliptic integrals.

\section{Resumming $\chi$ using its singularity structure}
\label{sec:chilog}

The origin of the logarithmic behavior in $(j - j_0)$ near the critical angular momentum $j_0$ is expected to be general (at least within our potential-based framework). This logarithmic behavior is simply related to the fact that the 
two largest positive real roots of the equation $p_\infty^2 - j^2 \bar{u}^2 + w(\bar{u},\gamma) = 0$ coalesce when $j \rightarrow j_0$.
Indeed, denoting these roots as $a(j)$ and $b(j)$, with $a(j) < b(j)$ when $j > j_0$ (e.g. $a=\bar{u}_2$ and $b=\bar{u}_3$ at 3PM), near the coalescence $a(j_0)=b(j_0)$, the scattering angle is given by an integral of the form 
\begin{equation}
	\chi \approx 2 j \int_{}^{a} \frac{d\bar{u}}{\sqrt{k(\bar{u}-a)(\bar{u}-b)}} \approx \frac{4 j}{\sqrt{k}} \ln \left[\frac{2 \sqrt{a}}{\sqrt{b-a}}\right]\,.
\end{equation}
In turn, this singularity implies a logarithmic singularity in $j$ near $j=j_0$ of the type 
\begin{equation}
	\label{eq:div_chi}
	\chi(j) \overset{\hspace{0.2cm} j \rightarrow j_0^+}{\sim} \frac{j}{j_0} \ln \left[\frac{1}{1-\frac{j_0}{j}}\right]\,.
\end{equation}

The existence of such universal logarithmic divergence suggests a procedure to resum the PM-expanded angle $\chi_{n{\rm PM}}(\gamma,j)$ by incorporating such a divergence. 
Specifically, we propose the following resummation procedure of $\chi_{n{\rm PM}}(\gamma,j)$.

For convenience, let us define the function 
\begin{equation}
	\label{eq:L}
	\mathcal{L}\left(x\right) \equiv \frac{1}{x} \ln \left[\frac{1}{1-x}\right]\,,
\end{equation}
such that, when $|x| < 1$, it admits the convergent power-series expansion 
\begin{equation}
	\mathcal{L}\left(x\right) = 1 + \frac{x}{2} + \frac{x^2}{3} + \cdots + \frac{x^{n-1}}{n} + \cdots\,.
\end{equation}
The logarithmic divergence in $\chi(j)$ near $j = j_0$, Eq.~\eqref{eq:div_chi}, precisely features the factor $\mathcal{L}\left(\frac{j_0}{j}\right)$.

We can now define the following $\mathcal{L}$-resummation of the PM-expanded $\chi_{n{\rm PM}} \sim 1/j + \cdots 1/j^n$ as being the unique function $\chi_{n{\rm PM}}^{\mathcal{L}}(j)$ of the form
\begin{equation}
	\label{eq:chilogPM}
	\chi_{n{\rm PM}}^{\mathcal{L}}(\gamma,j) = \mathcal{L}\left(\frac{j_0}{j}\right)\hat{\chi}_{n{\rm PM}}(\gamma,j;j_0)\,,
\end{equation}
where $\hat{\chi}_{n{\rm PM}}(\gamma,j;j_0)$ is a $n^{\rm th}$ order polynomial in $1/j$, say,
\begin{equation}
	\label{eq:hatchi}
	\hat{\chi}_{n{\rm PM}}(\gamma,j;j_0) \equiv \sum_{i = 1}^{n} 2 \frac{\hat{\chi}_i(\gamma;j_0)}{j^i}\,,
\end{equation}
such that the $n^{\rm th}$ first terms in the large-$j$ expansion of $\chi_{n{\rm PM}}^{\mathcal{L}}(\gamma,j)$ coincide with $\chi_{n{\rm PM}}(\gamma,j)$.
The latter condition uniquely determines the expressions of the $\hat{\chi}_i(\gamma;j_0)$ coefficients in terms of $j_0$ and of the original PM coefficients $\chi_i(\gamma)$, namely
\begin{align}
		\label{eq:hatchi_n}
\hat{\chi}_1 &= \chi_1\,, \nonumber \\
\hat{\chi}_2 &= \chi_2 - \frac{j_0}{2}\chi_1\,, \nonumber \\
\hat{\chi}_3 &= \chi_3 - \frac{j_0}{2}\chi_2 - \frac{j_0^2}{12}\chi_1\,, \nonumber \\ 
\hat{\chi}_4 &= \chi_4 - \frac{j_0}{2}\chi_3 - \frac{j_0^2}{12}\chi_2 - \frac{j_0^3}{24}\chi_1\,.
\end{align}

In order to define such a resummation procedure, we need to choose a value of $j_0$.
Such a value can be \textit{analytically} defined by comparing the successive terms of the expansion
\begin{equation}
	\mathcal{L}\left(\frac{j_0}{j}\right) = 1 + \frac{j_0}{2\,j} + \frac{j_0^2}{3\,j^2} + \cdots + \frac{j_0^{n-1}}{n\,j^{n-1}} + \cdots \,,
\end{equation} 
to the corresponding terms in 
\begin{equation}
	\frac{\chi_{n{\rm PM}}}{\chi_{1{\rm PM}}} = 1 + \frac{\chi_2}{\chi_1 \, j} + \frac{\chi_3}{\chi_1\, j^2} + \cdots + \frac{\chi_{n}}{\chi_1\, j^{n-1}}\,.
\end{equation} 

Following \textit{Cauchy's rule}, one expects to have a more accurate estimate of $j_0$ by using the highest known term in the PM expansion, i.e.
\begin{equation}
	\label{eq:j0PM}
j_0^{n{\rm PM}}(\gamma) \equiv \left[n\frac{\chi_n(\gamma)}{\chi_1(\gamma)}\right]^{\frac{1}{n-1}}\,, \hspace{1cm} n>1\,.
\end{equation}

In the present paper, we use this resummation procedure for two applications:
(i) for resumming $\chi_{\rm 4PM}$; and (ii) for fitting the NR data.

\section{Comparison between PM and NR scattering angles}
\label{sec:nr_comp}

In this section we compare four different analytical definitions of the scattering angles against NR simulations:
(i) the non-resummed PM-expanded scattering angle $\chi_{n{\rm PM}}$;
(ii) the $\mathcal{L}$-resummed PM scattering angle $\chi^{\mathcal{L}}_{n{\rm PM}}$;
(iii) the $w^{\rm eob}$-resummed PM scattering angle $\chi^{w\,{\rm eob}}_{n{\rm PM}}$;
and, finally, (iv) a sequence of scattering angles predicted by one specific EOBNR 
dynamics (\TEOBResumS{}~\cite{Nagar:2021xnh,Hopper:2022rwo}).

Concerning numerical results, we use one of the few publicly available NR simulation suites of BH scatterings~\cite{Damour:2014afa}.
Reference~\cite{Damour:2014afa} computed a sequence of ten simulations of equal-mass nonspinning BH binaries with (almost)\footnote{The initial energies of the NR simulations range between $\hat{E}_{\rm in} = 1.0225555(50)$ and $\hat{E}_{\rm in} = 1.0225938(50)$.	In all the plots, we fixed $\hat{E}_{\rm in}$ to the average value $\hat{E}_{\rm in} = 1.0225846$.} fixed initial energy $\hat{E}_{\rm in} \equiv E_{\rm in}/M \simeq 1.02258$ and with initial angular momenta $\hat{J}_{\rm in} \equiv J_{\rm in}/M^2$ varying between $\hat{J}_{\rm in} = 1.832883(58)$ (corresponding to an NR impact parameter $b_{\rm NR} = 16.0 M$) and $\hat{J}_{\rm in} = 1.099652(36)$ (corresponding to $b_{\rm NR} = 9.6 M$). 
The simulation corresponding to the latter impact parameter probes the strong-field interaction of two BHs and led to 
a scattering angle $\chi_{\rm NR} = 5.337(45)$ radians, i.e. at the beginning of the zoom-whirl regime.

\subsection{Comparing NR data to non-resummed ($\chi_{n{\rm PM}}$) and $\mathcal{L}$-resummed ($\chi^{\mathcal{L}}_{n{\rm PM}}$) scattering angles}

\begin{figure}[t]
	\includegraphics[width=0.48\textwidth]{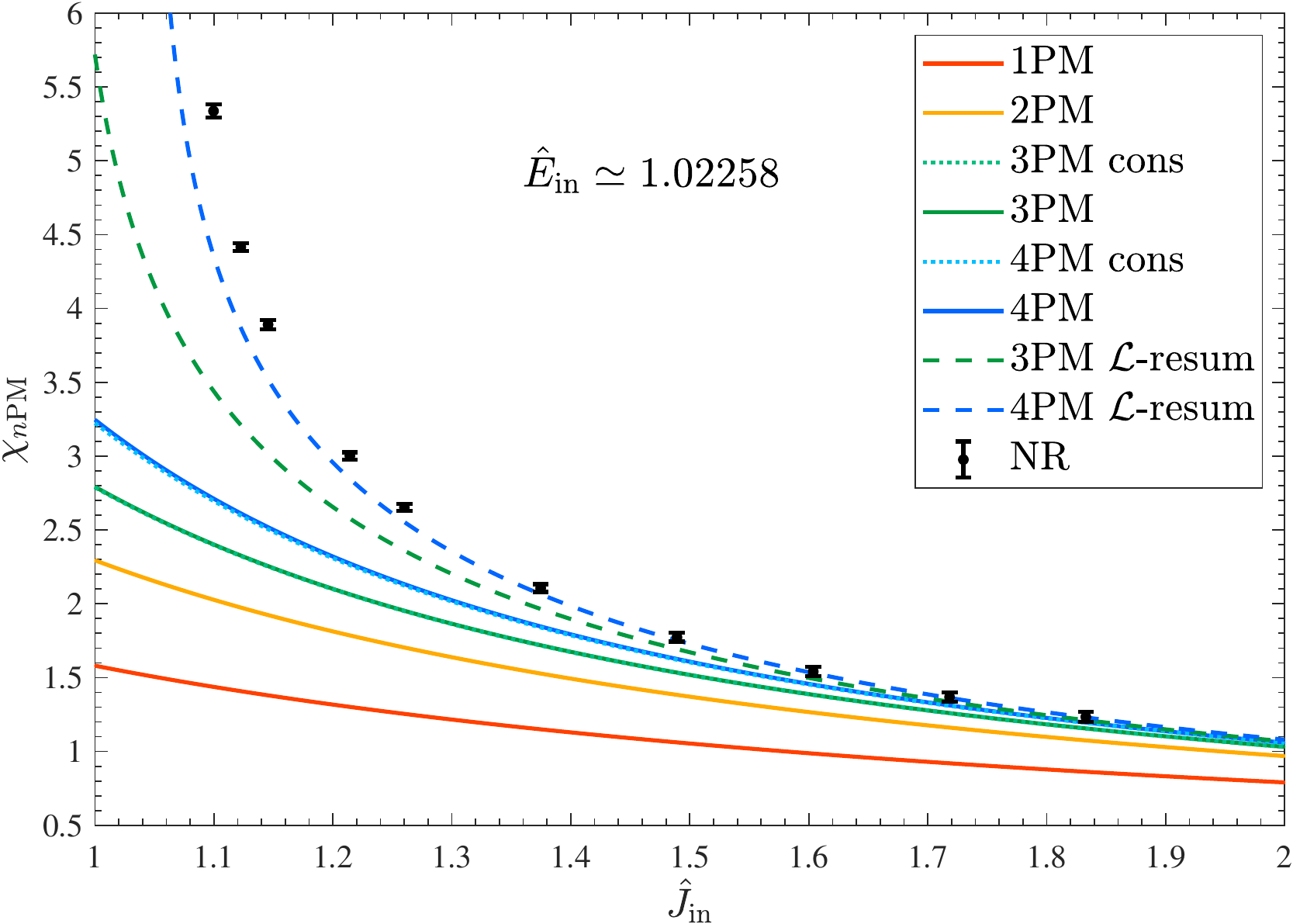}
	\caption{
		\label{fig:chiPM}
		Scattering angle comparison between the numerical results of Ref.~\cite{Damour:2014afa} and the PM-expanded scattering angles $\chi_{n{\rm PM}}$.
		In order to show the effect of radiative terms, we also plot the conservative part of the 3PM and 4PM scattering angles (dotted lines).
		The two topmost dashed lines represent the $\mathcal{L}$-resummed 3PN and 4PM (radiation-reacted) scattering angles $\chi^\mathcal{L}_{\rm 3PM}$ and $\chi^\mathcal{L}_{\rm 4PM}$.
	}
\end{figure}

In Fig.~\ref{fig:chiPM} we contrast NR data (indicated by black dots together with their error bars) to the sequence of analytically available PM-expanded scattering angles $\chi_{n{\rm PM}}$ (i.e. $1 \leq n \leq 4$), given by Eq.~\eqref{eq:chiPM}.
For the 3PM and 4PM orders we include both the conservative scattering angle and the radiation-reacted one.
This figure extends Fig.~6 of Ref.~\cite{Khalil:2022ylj} in two ways: 
(i) we include in the comparison the three smallest impact parameter simulations;
and (ii) we take advantage of recent analytical work~\cite{Manohar:2022dea,Dlapa:2022lmu,Bini:2022enm} 
to include radiation-reaction effects in the 4PM scattering.

As expected, PM expansions correctly capture the NR behavior for the largest values of the angular momentum, i.e. in the weak-field regime.
By contrast, for lower angular momenta (stronger fields), the differences between NR and PM results become large. 
Increasing the PM order improves the agreement with NR but is not enough to reach a satisfying agreement.
For instance, for the lower angular momentum datum, the 4PM-expanded (radiation-reacted) 
prediction yields $\chi_{\rm 4PM} \simeq 2.713$, while the NR result, $\chi_{\rm NR} = 5.337(45)$, is almost twice bigger.

In Fig.~\ref{fig:chiPM}, we added the (purely analytical) $\mathcal{L}$-resummed predictions for 
the radiation-reacted 3PM and 4PM angles (topmost dashed curves), using Eq.~\eqref{eq:chilogPM}, 
and the corresponding $j_0^{n{\rm PM}}$ from Eq.~\eqref{eq:j0PM}.
For clarity, we do not exhibit the corresponding conservative 3PM and 4PM scattering angles. 
They are both quite close to (but somewhat below) their respective radiation-reacted versions.
Figure~\ref{fig:chiPM} shows how the $\mathcal{L}$-resummation is quite efficient at improving the 
agreement between analytical PM information and NR data. For example, for the lower angular momentum 
datum, the $\mathcal{L}$-resummed 4PM (radiation-reacted) prediction is $\chi^\mathcal{L}_{\rm 4PM} \simeq 4.088$, 
which is only $\simeq 23\%$ smaller than the corresponding NR value.

\subsection{Comparing NR data to $w^{\rm eob}$-resummed ($\chi^{w\,{\rm eob}}_{n{\rm PM}}$) scattering angles}

\begin{figure}[t]
	\includegraphics[width=0.485\textwidth]{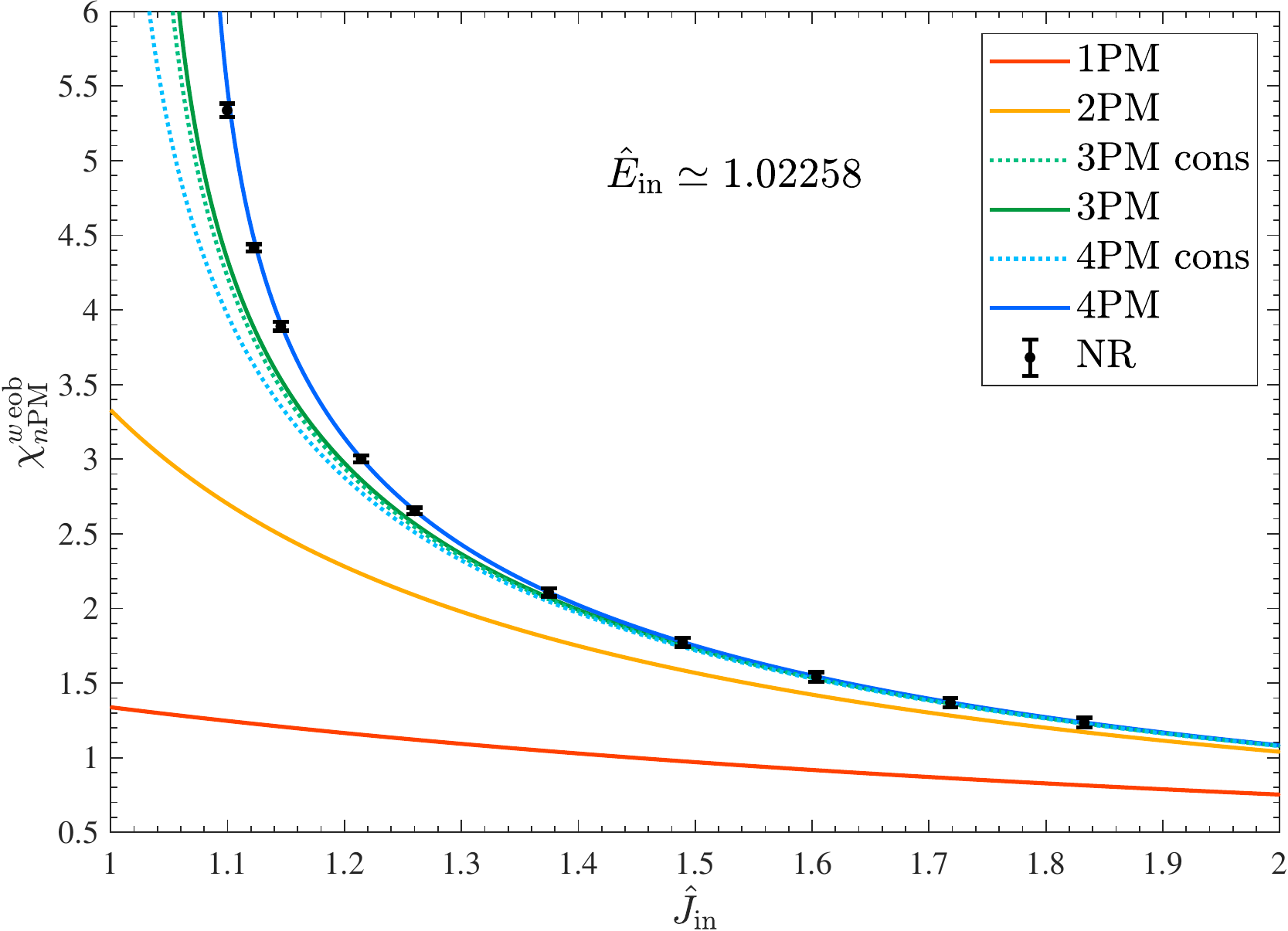}
	\caption{
		\label{fig:chiwPM2}
		Same comparison as Fig.~\ref{fig:chiPM} using the $w^{\rm eob}$-resummed scattering angles $\chi^{w\,{\rm eob}}_{n{\rm PM}}$ derived through the use of the EOB radial potentials $w_{n{\rm PM}}$.
		The agreement using 4PM results including radiation-reaction terms is excellent.
	}
\end{figure}

Fig.~\ref{fig:chiwPM2} compares the NR data to the $w^{\rm eob}$-resummed angles
$\chi^{w\,{\rm eob}}_{n{\rm PM}}$ [Eq.~\eqref{eq:chi_wPM}], i.e. to the sequence of angles computed by studying the scattering of a particle in the corresponding $n$PM-order potential $w_{n{\rm PM}}(\bar{r},\gamma) \sim 1/\bar{r} + \cdots + 1/\bar{r}^n$ [Eq.~\eqref{eq:wPM}].
Let us emphasize that each such potential is completely analytically defined from the corresponding $n$PM scattering angles, via Eq.~\eqref{eq:chi_to_w}.
Again, for the 3PM and 4PM orders, we show scattering angles computed through both the corresponding conservative and radiation-reacted $w(\bar{r},\gamma)$ potentials.

Apart from the conservative 4PM (light-blue dotted) curve, the $w^{\rm eob}$-resummed angles $\chi^{w\,{\rm eob}}_{n{\rm PM}}$ succeed in defining a sequence of approximants which not only gets closer to NR data as the PM order increases, but 
also reaches, at the (radiation-reacted) 4PM level, an excellent agreement with NR data. E.g., for the lower angular momentum datum, its prediction yields $\chi^{w\,{\rm eob}}_{\rm 4PM} \simeq 5.490$, which is only $\simeq 2.9\%$ higher than the corresponding NR result, $\chi_{\rm NR} = 5.337(45)$. 
This excellent agreement  is partly due to the fact that the critical angular momentum $j_0$ determined from the radial potential $w_{4{\rm PM}}(\bar{r},\gamma)$ [namely $j_0^{w_{4{\rm PM}}}(\gamma \simeq 1.09136) \simeq 4.3138$], is very close to the one obtained by $\mathcal{L}$-fitting the NR data [namely $j_0^{\rm fit} \simeq 4.3092 $].

The relatively poor performance of the $\chi^{w\,{\rm eob}}_{\rm 4PM, cons}$ prediction (which is slightly worse than both $\chi^{w\,{\rm eob}}_{\rm 3PM}$ and $\chi^{w\,{\rm eob}}_{\rm 3PM, cons}$) is rooted in the exceptional character of the conservative 4PM radial potential discussed above.
Indeed, the presence of a repulsive core in $w_{\rm 4PM}^{\rm cons}$ has a significant effect on the value of the critical $j_0$.
The corresponding critical angular momentum (at our considered energy $\gamma \simeq 1.09136$) is $j_0^{w_{\rm 4PM, cons}} \simeq 3.9983$, which is significantly below our other relevant estimates, namely: (i) $j_0^{w_{4{\rm PM}}} \simeq 4.3138$; (ii) $j_0^{w_{\rm 3PM}} \simeq 4.1432$; (iii) $j_0^{w_{\rm 3PM, cons}} \simeq 4.1198$; as well as (iv) $j_0^{\rm fit} \simeq 4.3092 $.
Let us also note that the effect of the second-order radiation-reaction contribution to $w_{\rm 4PM}(\bar{r},\gamma)$ is subdominant (at least for our parameter values) with respect to the first-order radiation-reaction one. To wit, the critical $j_0$ obtained by neglecting $w_4^{\rm rr,even}(\gamma)$ in $w_4(\gamma)$ is $j_0\simeq 4.3132$, which is very close to the complete one.
See next Section for more details.

\subsection{Comparing NR data to \TEOBResumS{} scattering angles}

\begin{figure}[t]
	\includegraphics[width=0.48\textwidth]{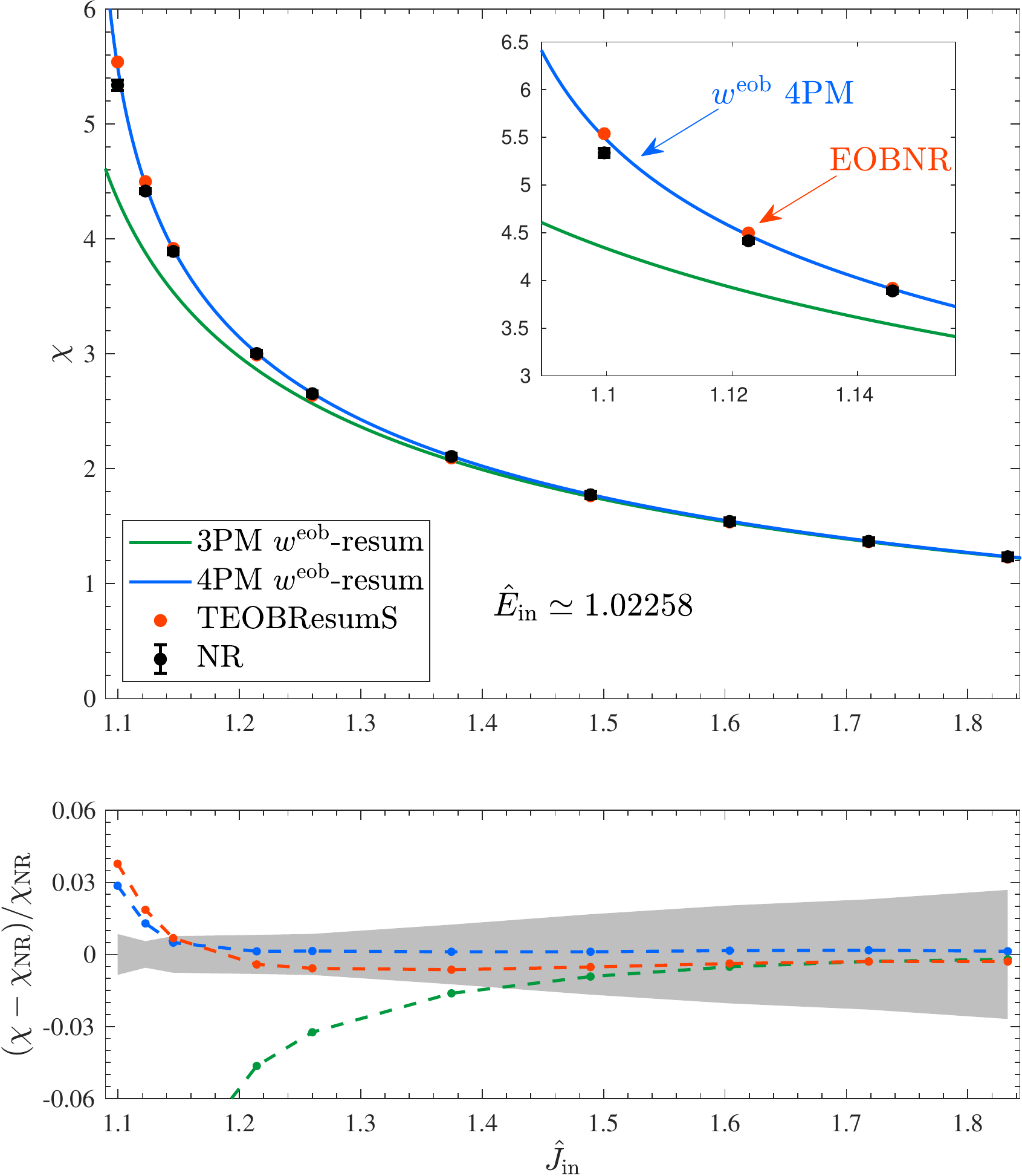}
	\caption{
		\label{fig:chiPM_TEOB}
		Comparison between NR simulations, PM results and the EOBNR model \TEOBResumS{}.
		Top panel: scattering angles. Bottom panel: fractional differences with respect to numerical results. The shaded grey area represents the NR errors.
	}
\end{figure}

Finally, in Fig.~\ref{fig:chiPM_TEOB}, we compare the scattering angle predictions of the EOBNR model \TEOBResumS{}\footnote{This EOB model combines an NR-calibrated high PN accuracy Hamiltonian with an analytical radiation-reaction force.}~\cite{Nagar:2021xnh,Hopper:2022rwo} to NR data and to the (radiation-reacted) $w^{\rm eob}$-resummed 3PM and 4PM scattering angles. This EOB model is run with initial conditions $(\hat{E}_{\rm in},\hat{J}_{\rm in})$.
The scattering angle predicted by \TEOBResumS{} (denoted by $\chi^ {\rm EOBNR}$) exhibits an excellent agreement with NR data for all angular momenta.
However, the bottom panel of Fig.~\ref{fig:chiPM_TEOB}, which displays the fractional differences with NR data (see corresponding Table~\ref{tab:chi_NR}), shows that the EOBNR differences are systematically (slightly) larger, in absolute value, than the $w_{\rm 4PM}$-NR ones.
Both of them, however, are compatible with the error bar on NR data, except for the two smallest impact-parameter data points. [The c.m. impact parameters, $b = \frac{h j}{p_\infty}$ of the first two rows of Table~\ref{tab:chi_NR} are respectively $b \simeq 10.20$ and $ b \simeq 10.41$.]

In order to further probe the relative performances of our two best 
analytical scattering predictions, we report in Table~\ref{tab:chi_Seth} the comparisons of $\chi^{\rm EOBNR}$ and $\chi^{w\,{\rm eob}}_{\rm 4PM}$ to the recent NR results of Ref.~\cite{Hopper:2022rwo}.
The latter simulations are somewhat complementary to the ones of Ref.~\cite{Damour:2014afa}, because they were performed with (almost) fixed angular momentum and varying energies.
Table~\ref{tab:chi_Seth} confirms the excellent performances of both \TEOBResumS{} and the $w^{\rm eob}$-resummation of the 4PM scattering angle.
Again, the EOBNR differences with respect to the numerical data are systematically larger (in absolute value) than the $w_{\rm 4PM}$ ones, though only the two smallest impact-parameter data points exhibit differences larger than the NR error bar. [The c.m. impact parameters of the last two rows of Table~\ref{tab:chi_Seth} are respectively $b \simeq 10.24$ and $ b \simeq 8.63$.]

\begin{table}[t]
	\caption{
		\label{tab:chi_NR} 
		Comparison between NR, \TEOBResumS{} and $w^{\rm eob}$-resummed 4PM scattering angles for the equal-mass, nonspinning configurations of Ref.~\cite{Damour:2014afa}.
		We report (in order): initial energy $\hat{E}_{\rm in}$; initial angular momentum $\hat{J}_{\rm in}$; NR scattering angle $\chi^{\rm NR}$; NR percentage error $\widehat{\sigma}\chi^{\rm NR}$; EOBNR scattering angle $\chi^{\rm EOBNR}$ and corresponding fractional difference with respect to NR data $\widehat{\Delta} \chi^{\rm EOBNR} \equiv \chi^{\rm EOBNR}/\chi^{\rm NR} -1$; $\chi_{\rm 4PM}^{w\,{\rm eob}}$ and fractional difference $\widehat{\Delta} \chi_{\rm 4PM}^{w\,{\rm eob}}$.
	}
	\begin{center}
		\begin{ruledtabular}
			\begin{tabular}{ c c | c c | c c | c c } 
				$\hat{E}_{\rm in}$ & $\hat{J}_{\rm in}$ & $\chi^{\rm NR}$ & $\widehat{\sigma} \chi^{\rm NR}$ & $\chi^{\rm EOBNR}$ & $\widehat{\Delta} \chi^{\rm EOBNR}$ & $\chi_{\rm 4PM}^{w\,{\rm eob}}$ & $\widehat{\Delta} \chi_{\rm 4PM}^{w\,{\rm eob}}$ \\
				\hline
				1.023 & 1.100 & 5.337 & 0.85\% & 5.539 &    3.78\% & 5.490 & 2.86\% \\
				1.023 & 1.123 & 4.416 & 0.55\% & 4.498 &    1.86\% & 4.473 & 1.29\% \\ 
				1.023 & 1.146 & 3.890 & 0.76\% & 3.917 &    0.68\% & 3.909 & 0.49\% \\ 
				1.023 & 1.214 & 3.002 & 0.81\% & 2.989 & $-$0.42\% & 3.006 & 0.13\% \\ 
				1.023 & 1.260 & 2.653 & 0.86\% & 2.638 & $-$0.58\% & 2.657 & 0.14\% \\ 
				1.023 & 1.375 & 2.107 & 1.24\% & 2.093 & $-$0.64\% & 2.109 & 0.11\% \\ 
				1.023 & 1.489 & 1.773 & 1.67\% & 1.764 & $-$0.53\% & 1.775 & 0.11\% \\ 
				1.023 & 1.604 & 1.541 & 2.04\% & 1.535 & $-$0.38\% & 1.544 & 0.16\% \\ 
				1.023 & 1.718 & 1.368 & 2.30\% & 1.364 & $-$0.30\% & 1.371 & 0.17\% \\ 
				1.023 & 1.833 & 1.234 & 2.69\% & 1.230 & $-$0.29\% & 1.236 & 0.13\% 
			\end{tabular}
		\end{ruledtabular}
	\end{center}
\end{table}

\begin{table}[t]
	\caption{
		\label{tab:chi_Seth} 
		Comparison between NR, \TEOBResumS{} and $w^{\rm eob}$-resummed 4PM scattering angles for the equal-mass, nonspinning configurations of Ref.~\cite{Hopper:2022rwo}.	The initial angular momentum is approximately constant, while the energy varies.
		The reported columns are the same of Table~\ref{tab:chi_NR}.
	}
	\begin{center}
		\begin{ruledtabular}
			\begin{tabular}{ c c | c c | c c | c c } 
				$\hat{E}_{\rm in}$ & $\hat{J}_{\rm in}$ & $\chi^{\rm NR}$ & $\widehat{\sigma} \chi^{\rm NR}$ & $\chi^{\rm EOBNR}$ & $\widehat{\Delta} \chi^{\rm EOBNR}$ & $\chi_{\rm 4PM}^{w\,{\rm eob}}$ & $\widehat{\Delta} \chi_{\rm 4PM}^{w\,{\rm eob}}$ \\
				\hline
				1.005 & 1.152 & 3.524 & 2.37\% & 3.500 & $-$0.69\% & 3.538 & 0.39\% \\ 
				1.015 & 1.152 & 3.420 & 0.66\% & 3.395 & $-$0.71\% & 3.421 & 0.04\% \\ 
				1.020 & 1.152 & 3.613 & 0.48\% & 3.614 & 0.02\%  & 3.625 & 0.32\% \\ 
				1.025 & 1.152 & 3.936 & 0.39\% & 3.997 & 1.54\%  & 3.977 & 1.02\% \\ 
				1.035 & 1.152 & 5.360 & 0.29\% & 6.038 & 12.63\% & 5.834 & 8.84\% \\ 
			\end{tabular}
		\end{ruledtabular}
	\end{center}
\end{table}

\section{Extracting the effective-one-body radial potential from numerical scattering data}
\label{sec:inversion}

In Secs.~\ref{sec:scatt_angle} and ~\ref{sec:chiw}, we have shown how to compute the scattering angle of a BH binary from the knowledge of a radial potential $w(\bar{r},\gamma)$, which encapsulates the general relativistic gravitational interaction in the presently used EOB formalism.
Let us now do the inverse: starting from the knowledge of a sequence of scattering angles at fixed energy (and varying $j$), we wish to extract the value of a corresponding (energy-dependent, radiation-reacted) radial potential $w_{\rm NR}(\bar{r},\gamma)$.
We will do so in two steps: (i) we replace the discrete set of NR scattering angles (at fixed energy) by a continuous function of $j$;
and (ii) we use Firsov's inversion formula~\cite{Landau:1960mec} (see also Ref.~\cite{Kalin:2019rwq}) to invert Eq.~\eqref{eq:chi_w}. 

\subsection{$\mathcal{L}$-resummation of NR data}
For the first step, we adapt the $\mathcal{L}$-resummation technique of Sec.~\ref{sec:chilog}, to the discrete sequence of NR data.
Namely, we define a continuous function of $j$, $\chi_{\rm NR}^{\rm fit}(j)$, by least-square fitting the ten NR data points\footnote{In doing so, we neglect the fractionally small differences in initial energies and only consider the average energy $\hat{E}_{\rm in} = 1.0225846$.} to a function of $1/j$ incorporating the logarithmically singular function $\mathcal{L}\left(j_0/j\right)$, defined in Eq.~\eqref{eq:L}.
More precisely, we use a fitting template of the general form 
\begin{align}
	\label{eq:chi_fit_gen}
	&\chi^{\rm fit}_{\rm gen}(j;j_0,a_{n+1},\cdots,a_{n+k}) = \nonumber \\ &\mathcal{L}\left(\frac{j_0}{j}\right)\left[\hat{\chi}_{n{\rm PM}}(j;j_0) + 2 \frac{a_{n+1}}{j^{n+1}} + \dots + 2 \frac{a_{n+k}}{j^{n+k}} \right]\,,
\end{align}
where $\hat{\chi}_{n{\rm PM}}$ is the ($j_0$-dependent) polynomial in $1/j$ defined in Eqs.~\eqref{eq:hatchi} and \eqref{eq:hatchi_n}.

Given any fitting template of the form \eqref{eq:chi_fit_gen}, the procedure is to determine the values of the $k+1$ free parameters $(j_0,a_{n+1},\cdots,a_{n+k})$ by least-square fitting the discrete NR data.
Such a procedure depends on the choice of PM-order, namely $n$, that we are ready to assume as known. Here we shall take $n = 3$ in order to incorporate the correct large-$j$ in a minimal way.
This choice makes our results completely independent from the recently acquired 4PM knowledge. 
Concerning the choice of $k$, we used the minimal value that led to a reduced chi-squared smaller than one. We found, rather remarkably, that it was enough to take $k=1$ and that increasing $k$ clearly led to overfitting. 

In conclusion, we used as fitting template the function
\begin{equation}
	\label{eq:chifit}
	\chi_{\rm NR}^{\rm fit}(j) = \mathcal{L}\left(\frac{j_0}{j}\right)\left[\hat{\chi}_{\rm 3PM}(j;j_0) + 2 \frac{a_4}{j^4} \right],
\end{equation}
which depends only on two\footnote{When using templates incorporating either no singularities in $j$ (polynomials in $1/j$) or a different singularity (e.g. a simple pole), we found that one needed more parameters and that the resulting fits seemed less reliable.} fitting parameters: an effective value of the critical angular momentum $j_0$; and an effective value of the 4PM-level parameter $a_4$.

The best-fit parameters are found to be
\begin{align}
	\label{eq:fit_params}
	a_4^{\rm fit} &= 9.61 \pm 0.68\,, \nonumber \\
	j_0^{\rm fit} &= 4.3092 \pm 0.0018\,,
\end{align}
leading to a reduced chi-squared $\chi^2/(10-2) \simeq 0.096$, corresponding to ten data points and two degrees of freedom.

One should keep in mind that this representation is only valid for 
one value of the energy, namely $\hat{E}_{\rm in} \simeq 1.02258$, corresponding to $\gamma \simeq 1.09136$, and for angular momenta $\hat{J}_{\rm in} \gtrsim 1.100$.
This formula also assumes that we are in the equal-mass case, i.e. $\nu = \frac14$, so that $j = 4\hat{J}_{\rm in}$.

Since we decided not to include the full analytical knowledge at our disposal but to instead leave the 4PM-level coefficient as a free parameter, 
it is possible to compare the analytical (radiation-reacted) 4PM coefficient $\chi_4(\gamma)$ [Eq.~\eqref{eq:chi_i}] to its corresponding NR-fitted value, say $\chi_{4}^{\rm fit}$, obtained by extracting the coefficient of $1/j^4$ in the expansion of $\frac12\chi_{\rm NR}^{\rm fit}(j)$ as a power-series in $1/j$.
For the considered energy, we find $\chi_{4}^{\rm fit} \simeq 58.11$, while its analytical counterpart is $\chi_{4} (\gamma \simeq 1.09136) \simeq 63.33$.

\subsection{Comparing various estimates of the critical angular momentum $j_0$}
\label{subsec:j0}

In the previous subsection, we compared NR-extracted information about $\chi_4$ to its analytical value.
Similarly, one can compare NR-extracted and analytical estimates of the critical angular momentum $j_0$ that determines the boundary between scattering and plunge.

Above, we indicated several analytical ways of estimating $j_0$.
As explained in Sec.~\ref{sec:chilog}, an analytical value of $j_0$, say $j_0^{w_{n{\rm PM}}}$, is determined at each PM level (with $n\geq3$) by studying the coalescence of the 
two largest positive real roots of the equation $\mathcal{E}_{(j,\gamma)}(\bar{u}) \equiv p_\infty^2 - j^2 \bar{u}^2 + w_{n{\rm PM}}(\bar{u},\gamma) = 0$.
As $w_{n{\rm PM}}(\bar{u},\gamma)$ is a polynomial in $\bar{u}$, the search of the critical $j_0$ is obtained by solving polynomial equations [discriminant of $\mathcal{E}_{(j,\gamma)}(\bar{u})$].

For the considered energy and equal masses, we find for the (radiation-reacted) 3PM and 4PM estimates
\begin{align}
j_0^{w_{3{\rm PM}}}(\gamma \simeq 1.09136) &\simeq 4.1432\,, \nonumber \\
j_0^{w_{4{\rm PM}}}(\gamma \simeq 1.09136) &\simeq 4.3138\,.
\end{align}
Note that the 4PM value is remarkably close to the fitting parameter $j_0^{\rm fit}$, Eq.~\eqref{eq:fit_params}.

For completeness, let us also mention the other (Cauchy-like) PM-related analytical estimates of $j_0$, say $j_0^{{\rm C}{n{\rm PM}}}(\gamma)$, defined in Eq.~\eqref{eq:j0PM}.
These are, at the (radiation-reacted) 3PM and 4PM levels
\begin{align}
	j_0^{\rm C3{\rm PM}}(\gamma \simeq 1.09136) &\simeq 3.8886\,, \nonumber \\
	j_0^{\rm C4{\rm PM}}(\gamma \simeq 1.09136) &\simeq 4.1890\,.
\end{align}
Note that the Cauchy-based 4PM value is rather close, though less close, to $j_0^{\rm fit}$, than the $w$-based estimate.

Besides the NR-based critical $j_0^{\rm fit}$ extracted here from the NR data of Ref.~\cite{Damour:2014afa}, other numerical simulations have estimated the value the critical value of $j_0$ in high-energy BH collisions~\cite{Shibata:2008rq,Sperhake:2009jz}. Reference~\cite{Shibata:2008rq} extracted the value of the critical impact parameter in the collision of equal-mass BHs with c.m. velocities $v_{\rm cm} = (0.6,0.7,0.8,0.9)$.
Reference~\cite{Sperhake:2009jz} extracted $j_0$ for $v_{\rm cm} = 0.94$.

\begin{figure}[t]
	\includegraphics[width=0.48\textwidth]{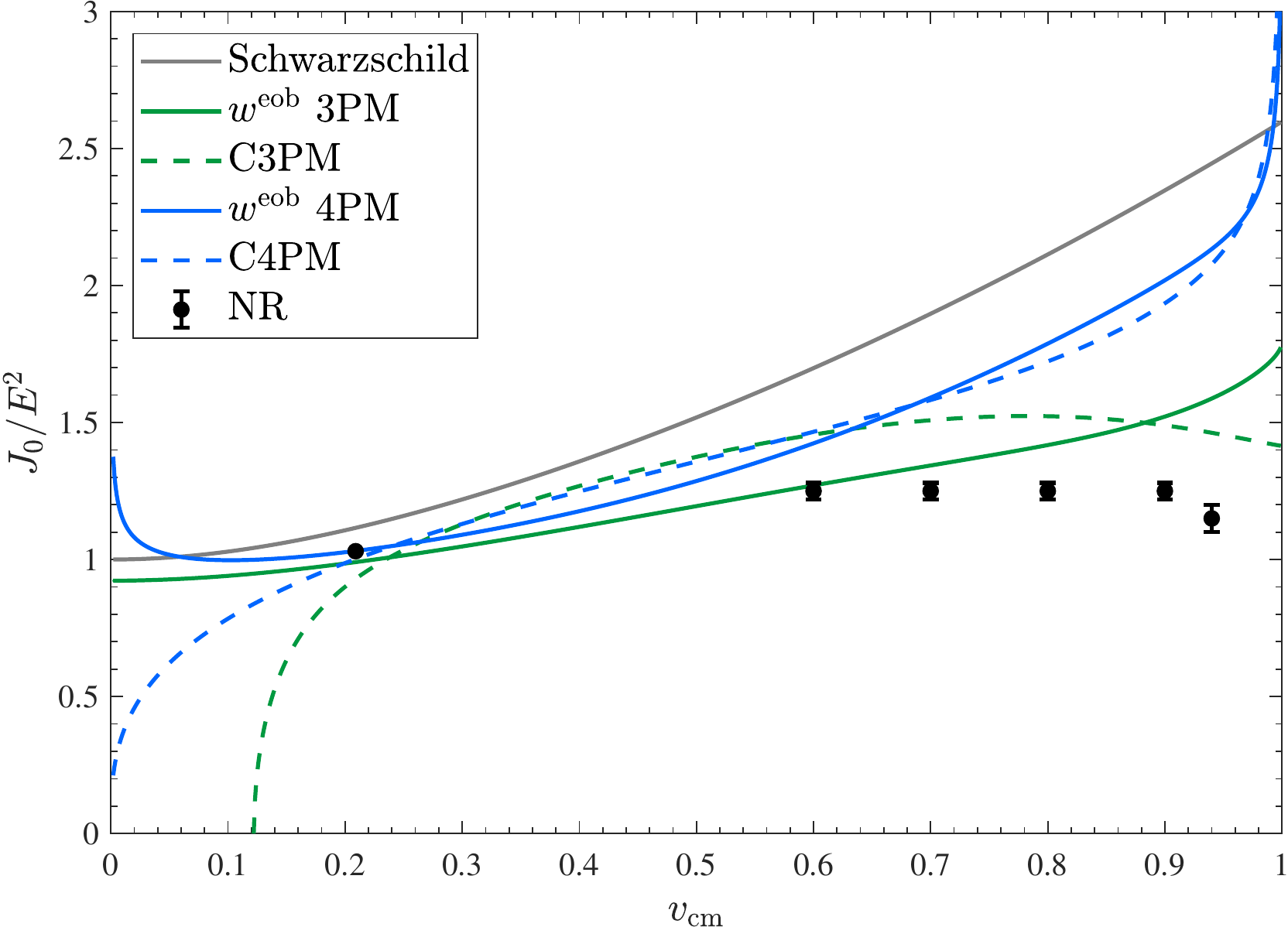}
	\caption{
		\label{fig:j0}
		Comparison between various analytical estimates of the critical (rescaled) angular momentum $J_0/E^2$ and NR results.
		We display the value determined by the (radiation-reacted) $w_{\rm 3PM}$ and $w_{\rm 4PM}$, together with the respective estimates using Cauchy's rule [Eq.\eqref{eq:j0PM}].
		The Schwarzschild estimate (grey line) is plotted as a reference.
		The NR points are from left to right: (i) our estimate coming from the fit of Eqs.~\eqref{eq:chifit} and \eqref{eq:fit_params};
		(ii) the four values computed in Ref.~\cite{Shibata:2008rq};
		(iii) the value calculated in Ref.~\cite{Sperhake:2009jz}.
	}
\end{figure}

It is convenient to express these results in terms of the quantity $J_0/E^2$ (measuring the dimensionless ``Kerr parameter'') of the system, namely
\begin{equation}
	\frac{J_0}{E^2} = \frac{\nu\, j_0}{1+2\nu\left(\gamma-1\right)}\,.
\end{equation}
In Fig.~\ref{fig:j0}, we compare various estimates of $J_0/E^2$ for the equal-mass case, $\nu = \frac{1}{4}$, as a function of the c.m. velocity 
\begin{equation}
	v_{\rm cm} = \sqrt{\frac{\gamma-1}{\gamma+1}}\,.
\end{equation}

The topmost curve in Fig.~\ref{fig:j0} is the test-mass (Schwarzschild) estimate of $J_0/E^2$, which is plotted as a reference.
It monotonically increases with $v_{\rm cm}$ from 1 at $v_{\rm cm} = 0$ to $\frac32\sqrt{3} \approx 2.598$ at $v_{\rm cm} = 1$.

The numerical simulations suggest that the critical $J_0^{\rm NR}/E^2$ has a finite limit, slightly higher than 1, in the ultra-high-energy regime $v_{\rm cm} \rightarrow 1$ ($\gamma \rightarrow \infty$). The last NR data point~\cite{Sperhake:2009jz} is $J_0^{\rm NR}/E^2 = (1.175 \pm 0.025)$ for $v_{\rm cm} = 0.94$, i.e. $\gamma \simeq 16.18$.

For the mildly-relativistic velocities considered in Ref.~\cite{Damour:2014afa}, $v_{\rm cm} \simeq 0.2090$ ($\gamma \simeq 1.09136$),  our $w_{\rm 4PM}$ estimate, $J_0^{w{\rm 4PM}}/E^2 \simeq 1.03134$, is the closest to the NR result, $J_0^{\rm NR, fit}/E^2 = (1.03024 \pm 0.00043)$.
By contrast, Fig.~\ref{fig:j0} suggests that both 4PM-level estimates become inaccurate for velocities $v_{\rm cm} \gtrsim 0.6$ (corresponding to $\gamma \gtrsim 2.125$). 
Actually, both 4PM-level estimates of $J_0/E^2$ have a power-law divergence when $v_{\rm cm} \rightarrow 1$ ($\gamma \rightarrow \infty$) linked to the power-law divergence of $\chi_{4}/\gamma^4 \propto \gamma^{1/2}$ in the high-energy limit~\cite{Bini:2022enm}.
We leave to future work a study of possible ways to cure the bad high-energy behaviour of $\chi_4(\gamma)$ and of its corresponding potentials.

On the other hand, both (radiation-reacted) 3PM-level analytical estimates seem to be in qualitative agreement with corresponding NR data, especially in the high-energy limit.
This is probably linked to the good high-energy behavior of the \textit{radiation-reacted} 3PM scattering angle~\cite{Amati:1990xe,Damour:2020tta,DiVecchia:2021ndb}.

Apart from the Schwarzschild and $w^{\rm eob}$ 3PM estimates, the other curves in Fig.~\ref{fig:j0} feature a bad behaviour in the low-velocity limit, $v_{\rm cm} \to 0$ ($\gamma \to 1$).
The $J_0/E^2$ estimate from $w_{\rm 4PM}$ blows up because the linear radiation reaction contribution to $w_4$ blows up like $w_{4}^{\rm rr,odd}(\gamma) \approx \frac{34 \sqrt{2} \nu}{9}(\gamma-1)^{-1/2}$ when $\gamma \to 1$.
The Cauchy-based 4PM estimate for $J_0/E^2$ vanishes proportionally to $(\gamma -1)^{1/6}$ in the low-velocity limit (because of the low-velocity blow-up of $\chi_1$).
Finally, the Cauchy-based 3PM estimate for $J_0/E^2$ has a branch-cut singularity $\propto (\gamma - \gamma_*)^{1/2}$ because $\chi_{3}(\gamma;\nu)$ changes sign at $\gamma_*(\nu)$ [with $\gamma_*(1/4) \simeq 1.0303$, corresponding to $v_{\rm cm}^* \simeq 0.1222$].

In order to clarify the physics of scattering BHs it would be important to fill the gaps in NR data visible in Fig.~\ref{fig:j0} by exploring a larger range of c.m. velocities.

\subsection{Extracting the EOB radial potential $w_{\rm NR}$ from NR data}
\label{subsec:inv}
Let us now come to the second step of our strategy for extracting information from NR data.
It is based on Firsov's inversion formula, which reads
\begin{equation}
\label{eq:Firsov_j}
\ln \left[1 + \frac{w(\bar{u},p_\infty)}{p_\infty^2}\right] = \frac{2}{\pi} \int_{\bar{r} |p(\bar{r},\gamma)|}^{\infty} dj \frac{\chi (\gamma, j)}{\sqrt{j^2 - \bar{r}^2 \, p^2(\bar{r},\gamma)}}\,,
\end{equation}
where 
\begin{equation}
	p^2(\bar{r},\gamma) \equiv p_\infty^2 + w(\bar{u},p_\infty)\,.
\end{equation}
Introducing the rescaled radial potential
\begin{equation}
\hat{w}(\bar{r},\gamma) \equiv \frac{w(\bar{r},\gamma)}{p_\infty^2}\,,
\end{equation}
and an \textit{effective} impact parameter
\begin{equation}
b \equiv \frac{j}{p_\infty},
\end{equation}
Eq. \eqref{eq:Firsov_j} becomes
\begin{equation}
	\label{eq:w_inv}
	\ln \left[1 + \hat{w}(\bar{r},\gamma)\right]
	= \frac{2}{\pi} \int_{\bar{r}\, \sqrt{1 + \hat{w}(\bar{r},\gamma)}}^{\infty} db \frac{\chi (\gamma, b)}{\sqrt{b^2 - \bar{r}^2 \left[1 + \hat{w}(\bar{r},\gamma)\right]}}\,.
\end{equation}

This is a recursive expression for defining $\hat{w}(\bar{r},\gamma)$ that can be solved iteratively.

Instead of solving Eq.~\eqref{eq:w_inv} iteratively to get $\hat{w}$ as a function of $\bar{r}$, we can obtain a parametric representation of both $\hat{w}$ and $\bar{r}$ as functions of an auxiliary parameter $\rho$ by defining the function
\begin{equation}
	\label{eq:Arho}
	A_{\chi}(\rho) \equiv \frac{2}{\pi} \int_{\rho}^{\infty} db \frac{\chi (\gamma, b)}{\sqrt{b^2 - \rho^2}}\,,
\end{equation}
which is related to the Abel transform of $\chi(\gamma,b)$.

In terms of the function $A_{\chi}(\rho)$, we get the exact parametric representation
\begin{align}
	\label{eq:w_par}
	\hat{w}(\rho) = -1 + e^{A_\chi(\rho)}\,, \nonumber \\
	\bar{r}(\rho) = \rho \, e^{- \frac12 A_\chi(\rho)}\,.
\end{align}

Note that the value of the parameter $\rho$, when considered as a function of $\bar{r}$, is
\begin{equation}
	\rho(\bar{r},\gamma) = \bar{r}\, |p(\bar{r},\gamma)| = \bar{r}\, \sqrt{1 + \hat{w}(\bar{r},\gamma)}\,.
\end{equation}

Equations~\eqref{eq:w_par} and \eqref{eq:Arho} allow us to extract information from the NR scattering data of Ref.~\cite{Damour:2014afa}.

Inserting our fit, Eqs.~\eqref{eq:chifit} and \eqref{eq:fit_params}, into Eqs.~\eqref{eq:w_par} and \eqref{eq:Arho}, we are able to numerically compute an NR-estimate of the (EOB) radial potential $w_{\rm NR}(\bar{r},\gamma)$.
The latter radial potential is determined from NR data only down to a radius $\bar{r}$ corresponding to the lowest $j$ for which NR simulations are available.
In our case, the minimum $\hat{J}_{\rm in} = 1.099652(36)$ corresponds to $\bar{r}_{\rm min} \simeq 2.567$.

\begin{figure}[t]
	\includegraphics[width=0.48\textwidth]{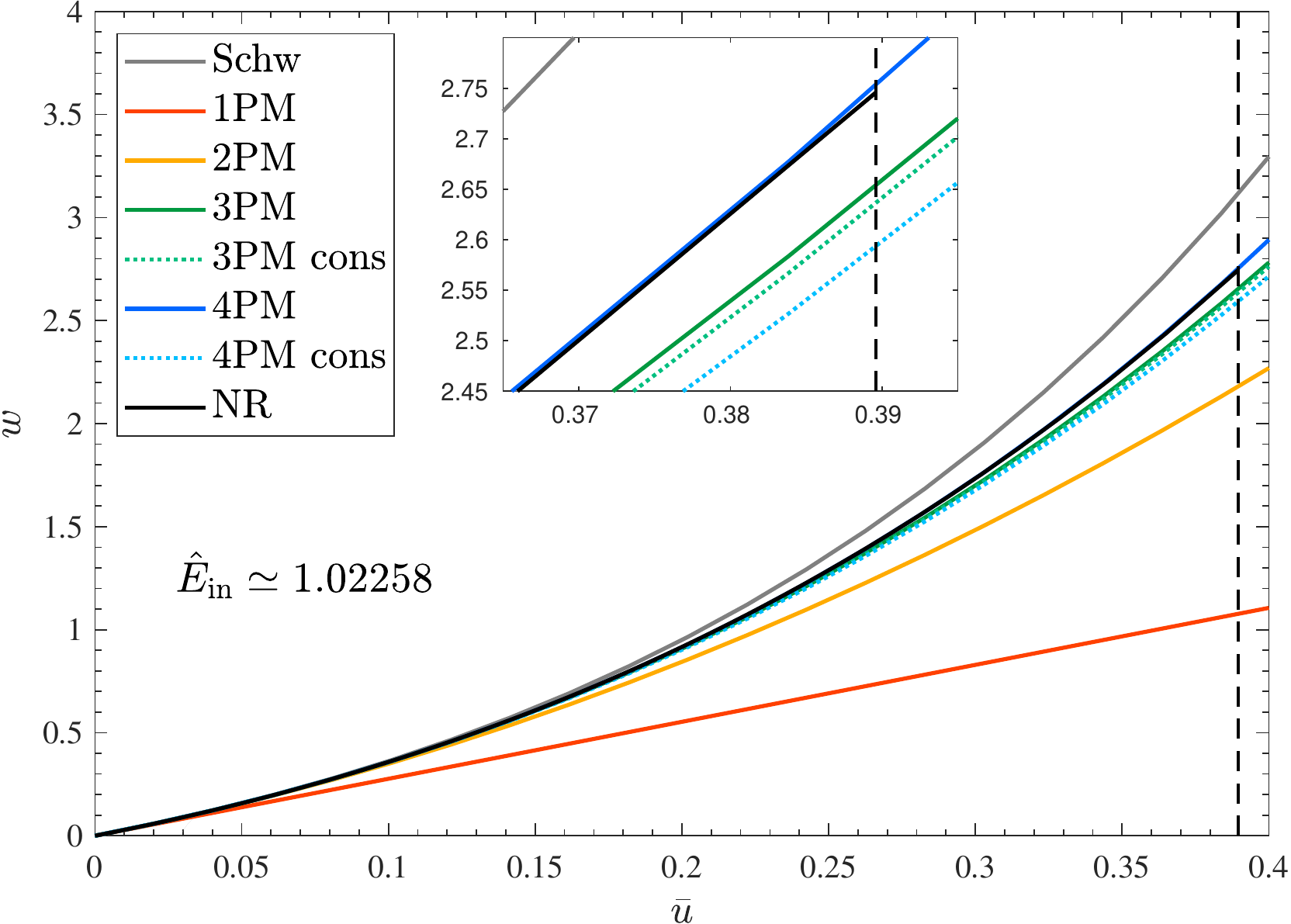}
	\caption{
		\label{fig:wNR}
		Comparison between the radial potential $w_{\rm NR}$, extracted from NR simulations up to $\bar{u}_{\rm max} = 1/\bar{r}_{\rm min} \simeq 0.3891$ (vertical dashed line), and the corresponding $w_{n{\rm PM}}$ ones.
		The series of PM-expanded potentials converges towards the NR-extracted one. In particular, the (radiation-reacted) 4PM potential is remarkably close to $w_{\rm NR}$.
		The test-mass potential, $w^{\rm Schw}$ (grey line), is plotted as a reference.
	}
\end{figure}

In Fig.~\ref{fig:wNR} we display the EOB-type (isotropic coordinates) radial potential extracted from the numerical data of Ref.~\cite{Damour:2014afa}, $w_{\rm NR}(\bar{r})$ (here plotted as a function of $\bar{u}\equiv1/\bar{r}$). 
We again remind the reader that this radial potential is energy-dependent and contains radiation-reaction effects.
It is determined here only for $\gamma \simeq 1.09136$.
Fig.~\ref{fig:wNR} compares $w_{\rm NR}(\bar{u})$ to the PM-expanded potentials $w_{n{\rm PM}}(\bar{u})$ for $1\leq n\leq4$.
For $n=3$ and 4, we exhibit both the conservative and the radiation-reacted avatars of the potential.
As we expected from the scattering angle comparison above, the \textit{radiation-reacted} 4PM potential is remarkably close to the NR one (see inset).
By contrast, the conservative 4PM potential is less close to $w_{\rm NR}(\bar{u})$ than the 3PM ones.
The error bar on $w_{\rm NR}(\bar{u})$ coming from NR inaccuracies, together with our fitting procedure, would be barely visible and will be discussed below.

\begin{figure}[t]
	\includegraphics[width=0.48\textwidth]{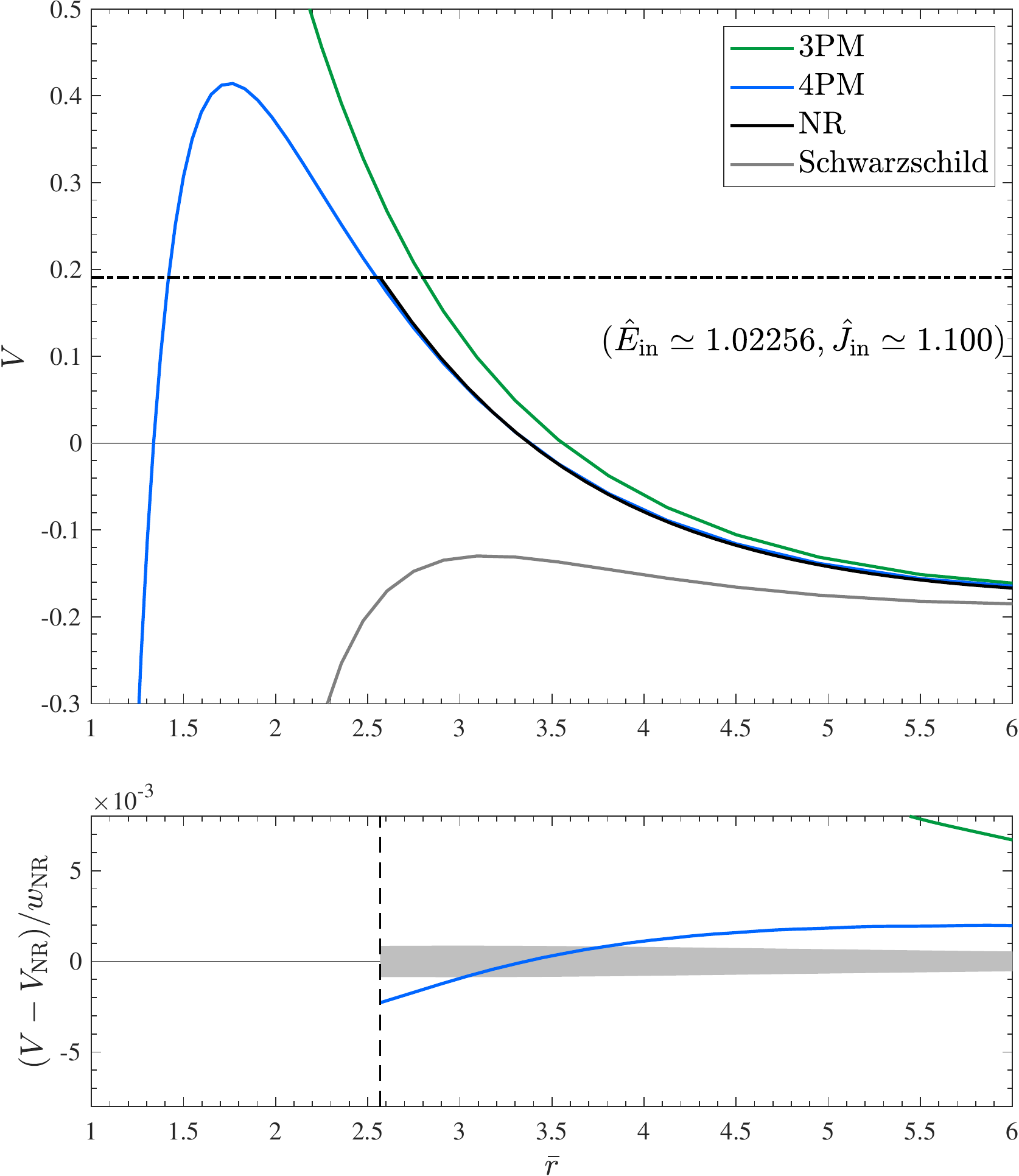}
	\caption{
		\label{fig:VNR}
		Top panel: comparison between the gravitational potential $V$ extracted from NR simulations and some of the PM ones shown in Fig.~\ref{fig:wPM2}.
		The horizontal line, corresponding to $p_\infty^2$, marks the maximum point up to which we can extract information from the numerical simulations.
		Bottom panel: fractional differences between PM potentials and the NR one, expressed as $(V_{n{\rm PM}}-V_{\rm NR})/w_{\rm NR} = (w_{\rm NR}-w_{n{\rm PM}})/w_{\rm NR}$. The shaded area is an estimated fractional error computed using the fit errors of Eq.\eqref{eq:chifit}.
		The (radiation-reacted) 4PM $w$ potential fractionally differs from $w_{\rm NR}$ only by $\sim \pm 2\times10^{-3}$.
	}
\end{figure}

For reference, we also displayed in Fig.~\ref{fig:wNR}, the (exact) Schwarzschild $w$ potential defined by setting $\hat{Q} = 0$ in Eqs.~\eqref{eq:w} and \eqref{eq:ASchw}.
Note that $w_{\rm NR}(\bar{r})$ lies significantly below $w_{\rm Schw}(\bar{r})$ which gives a direct NR-based proof that the Einsteinian gravitational interaction between two equal-mass BHs is \textit{less attractive} than its test-mass limit.

In order to extract the $w$ potential for other energies, and other mass ratios, one would need corresponding NR simulation suites with fixed energy, fixed mass ratio, and varying angular momentum.

A complementary view of the physics described by the potential $w_{\rm NR}(\bar{r})$ is presented in Fig.~\ref{fig:VNR}.
This figure contrasts several versions of the Newtonian-like potential $V(\bar{r},\gamma,j)$, defined in Eq.~\eqref{eq:V}, plotted versus $\bar{r}$.
We recall that this potential combines a (repulsive) centrifugal potential $j^2/\bar{r}^2$ with the (attractive) radial potential $-w(\bar{r},\gamma)$.
For definiteness, we use in Fig.~\ref{fig:VNR} the specific values of initial energy, $\hat{E}_{\rm in} \simeq 1.02256$, and angular momentum, $\hat{J}_{\rm in} = j/4 \simeq 1.100$, corresponding to the smallest impact-parameter simulation used here.

Figure~\ref{fig:VNR} compares $V_{\rm NR}(\bar{r},\hat{E}_{\rm in},\hat{J}_{\rm in})$, which is undefined below $\bar{r}_{\rm min} \approx 2.567$, to two PM-informed $V$ potentials: the (radiation-reacted) 3PM and 4PM ones. For reference, the test-mass limit potential $V^{\rm Schw}$ is also included.
 
This figure shows to what extent the smallest impact-parameter simulation has probed the topmost part of the $V$ potential.
Simulations with the same energy but smaller angular momenta, especially in the range between $\frac14j_0^{\rm fit} \simeq \frac14j_0^{w_{\rm 4PM}} \simeq 1.078$ and $\hat{J}_{\rm in}\simeq 1.100$, would allow one to explore the hill-top of the $V$ potential.
This figure also shows that the potential ruling the dynamics of equal-mass BH collisions is quite different from its test-mass limit.

The lower panel of Fig.~\ref{fig:VNR} displays the fractional differences between the considered PM-informed $V$ potentials and the NR one, expressed as $(V_{n{\rm PM}}-V_{\rm NR})/w_{\rm NR} = (w_{\rm NR}-w_{n{\rm PM}})/w_{\rm NR}$.
The shaded area is an estimate of the fractional error on $V_{\rm NR}$ computed by adding in quadrature the 68$\%$ confidence-level errors
on our (2-parameter) best-fit template, Eqs.~\eqref{eq:chifit} and \eqref{eq:fit_params}.
The (radiation-reacted) 4PM $w$ potential fractionally differs from $w_{\rm NR}$ only by $\sim \pm 2\times10^{-3}$.

\section{Conclusions}
\label{sec:concl}

In this work, we first reviewed the current knowledge of the PM scattering angle up to 4PM~\cite{Bern:2021dqo,Bern:2021yeh}, including radiative terms~\cite{Manohar:2022dea,Dlapa:2022lmu,Bini:2022enm}.
We then emphasized that, instead of using the PM results in the standard form of a PM-expanded scattering angle, the same PM information can be usefully reformulated in terms of radial potentials entering a simple EOB mass-shell condition~\cite{Damour:2017zjx}.
We found that this reformulation ($w^{\rm eob}$-resummation) greatly improves the agreement between the scattering angle obtained 
from PM analytical results and NR simulations of equal-mass, nonspinning BH binaries~\cite{Damour:2014afa,Hopper:2022rwo}. 
See Fig.~\ref{fig:chiwPM2}, to be compared to the four bottom curves of Fig.~\ref{fig:chiPM}.
The scattering angle computed using a radiation-reacted 4PM $w$ potential is as accurate (and even slightly more accurate) than 
the scattering angle computed by using one of the state-of-the-art (NR-calibrated, high-PN accuracy) EOB dynamics, 
namely the \TEOBResumS{} one~\cite{Nagar:2021xnh,Hopper:2022rwo}. 
The agreement between NR data and $\chi_{\rm 4PM}^{w\,{\rm eob}}$ is better than $1\%$ for most data points.
See Fig.~\ref{fig:chiPM_TEOB} and Tables~\ref{tab:chi_NR} and~\ref{tab:chi_Seth}.

Separately from the just mentioned $w^{\rm eob}$-resummation, we introduced a new resummation technique of scattering angles ($\mathcal{L}$-resummation).
This technique consists in incorporating a logarithmic singularity (corresponding to the critical angular momentum separating scattering motion from plunging ones) into the representation of the scattering angle as a function of angular momentum. 
We showed the usefulness of the $\mathcal{L}$-resummation technique in two specific applications:
(i) resummation of the PM-expanded scattering angles (see dashed lines in Fig.~\ref{fig:chiPM}, to be contrasted to the non-resummed bottom lines);
and (ii) accurate representation of a discrete sequence of NR scattering data by means an analytic fitting template, see Eqs.~\eqref{eq:chifit} and~\eqref{eq:fit_params}.

In Sec.~\ref{subsec:j0} we compared various estimates of the critical angular momentum $J_0$ separating scattering motions from coalescing ones.
This critical $J_0$ is a function of the initial energy and of the symmetric mass ratio. In Fig.~\ref{fig:j0} we compare four different PM-based analytic estimates of $J_0$ to NR data of both mildly-relativistic~\cite{Damour:2014afa} and highly-relativistic~\cite{Shibata:2008rq,Sperhake:2009jz} BH scatterings.
Figure~\ref{fig:j0} shows that the excellent agreement that we found between the radiation-reacted 4PM-based estimate of $j_0$ and the mildly relativistic NR data of Ref.~\cite{Damour:2014afa} ($v_{\rm cm} \simeq 0.2$) is not maintained at c.m. velocities $v_{\rm cm} \gtrsim 0.6$.
This is probably linked to the anomalous power-law behavior of $\chi_{\rm 4PM}$ in the high-energy limit.
This shows the need to numerically explore BH scattering in the range of velocities bridging the present gap between mildly-relativistic 
and highly-relativistic regimes, i.e. $0.2 \lesssim v_{\rm cm} \lesssim 0.6$.
The low-velocity regime $v_{\rm cm} \lesssim 0.2$ would also be quite interesting to explore.

Finally, we made use of Firsov's inversion formula to extract, for the first time, from NR data a gravitational 
potential, $w_{\rm NR}$, describing, within the EOB framework, the scattering of two BHs.
This potential contains both conservative and radiation reaction effects and is determined here only for the specific 
initial energy $\hat{E}_{\rm in}\simeq 1.02258$.  
We found that the (radiation-reacted) EOB potential $w_{\rm 4PM}$ is remarkably close to 
the NR-extracted one (see Figs.~\ref{fig:wNR} and~\ref{fig:VNR}).

This result opens a new avenue for extracting useful information from NR simulations of scattering BHs.
The computation of additional sequences of configurations at constant energy and varying angular momentum 
down to their critical $J_0$ will help to extend (following the strategy of Sec.~\ref{subsec:inv}) the knowledge of 
the energy-dependent radial potential $w_{\rm NR}(\bar{r},\gamma)$ to a larger range of energies.
This seems within reach of public codes like the Einstein Toolkit~\cite{Loffler:2011ay,EinsteinToolkit:2022_11}, which 
was already employed to obtain the NR scattering angles~\cite{Damour:2014afa,Hopper:2022rwo} used in this work. 

A parallel extension of both the procedure of Sec.~\ref{subsec:inv}, and of NR simulations, to 
spinning and unequal-mass systems is evidently called for. Such a knowledge will be a useful guideline to probe and 
compare the accuracy of various theoretical results (PM-based, PN-based EOBNR, $\dots$) and will offer new prospects 
for improving the accuracy of templates for eccentric and hyperbolic systems.

The unprecedented agreement between PM-based information and numerical results presented here was obtained 
by using one specific way (isotropic gauge, non-resummed energy-dependent $w$, $\dots$) of incorporating PM 
information into an EOB framework. 
We leave to future work an exploration of other ways of integrating PM information into EOB theory,
such as the use of a post-Schwarzschild framework, using Schwarzschild-type coordinates and the corresponding PM-expansion of the Finsler-type term $Q$ of Eq.~\eqref{eq:massshell},
or different ways of incorporating radiative effects (e.g., as in Ref.~\cite{Damour:2014afa}, or by adding a radiation reaction force). 

Finally, the NR-PM comparison presented here gives a new motivation to exploit the analytical flexibility of the EOB approach 
and to explore various ways of using PM information so as to improve our analytical 
description of binary systems not only in scattering situations, but also in bound states.

\section*{Acknowledgments}
P. R. thanks the hospitality and the stimulating environment of the Institut des Hautes Etudes Scientifiques. 
We thank A.~Nagar for collaboration at the beginning of this project, for suggestions and a careful reading
of the manuscript. 
The authors are also grateful to D.~Bini and R.~Russo for useful 
discussions during the development of this work and 
to O.~Long for pointing out some typos in the first version of this manuscript.
P.~R. is supported by the Italian Minister of University and Research (MUR) via the 
PRIN 2020KB33TP, {\it Multimessenger astronomy in the Einstein Telescope Era (METE)}.
The present research was also partly supported by the ``\textit{2021 Balzan Prize for 
Gravitation: Physical and Astrophysical Aspects}'', awarded to Thibault Damour.

\appendix

\bibliographystyle{apsrev4-1}
\bibliography{refs20230302.bib, local.bib}

\end{document}